%% file: main.tex
\documentclass[5p]{elsarticle}
\usepackage{amsmath}
	\numberwithin{equation}{section}

\usepackage[british]{babel}
\usepackage{tabularx}
\usepackage{multirow}
\usepackage{booktabs}
\usepackage{longtable}
\usepackage{rotating}
\usepackage{lineno}
\usepackage{pdflscape}
\usepackage[table]{xcolor}
\usepackage{subcaption}
\usepackage[export]{adjustbox}
\usepackage{amsmath} 
\usepackage{mathtools}

\usepackage{letltxmacro}

\usepackage{amssymb}
\usepackage{amsmath}
\newcommand{\angstrom}{\text{\normalfont\AA}}
\usepackage{amssymb}
\usepackage{amsthm}
\usepackage{mathtools}
\usepackage{braket}
\usepackage{bm}
\usepackage{bbm}
\usepackage{graphicx}
\usepackage{xcolor}
\usepackage{bbold}
\usepackage{blindtext}
\usepackage{graphicx}
\usepackage{graphics}
\usepackage{verbatim}   
\usepackage{amsfonts}
\usepackage{adjustbox}
\usepackage{mathrsfs}  

\usepackage{letltxmacro}

\usepackage{siunitx}

\usepackage[T1]{fontenc}
\usepackage[utf8]{inputenc}

\makeatletter
\newcommand\thefontsize{The current font size is: \f@size pt}
\makeatother

\usepackage{lineno,hyperref}
\modulolinenumbers[5]

\journal{Acta Materialia}

\begin{document}

\begin{frontmatter}


\title{\textcolor{black}{Atomistic simulation of martensite microstructural evolution during temperature driven $\beta \rightarrow \alpha$ transition in pure titanium}}

\author[onera,lspm,epfl]{C. Baruffi}
\author[onera]{A. Finel \fnref{myfootnote}}
\author[onera]{Y. Le Bouar}
\author[lspm]{B. Bacroix}
\author[lspm]{O.U. Salman}

\fntext[myfootnote]{alphonse.finel@onera.fr}

\address[onera]{Université Paris-Saclay, ONERA, CNRS, LEM, 29 Avenue de la Division Leclerc, 92320 Ch\^{a}tillon, France}
\address[lspm]{CNRS, LSPM UPR340, Université Sorbonne Paris Nord, 93430 Villetaneuse, France}
\address[epfl]{Ecole polytechnique fédérale de Lausanne (EPFL), Faculté STI, LAMMM, Lausanne CH-1015, Switzerland}


\begin{abstract}
Titanium and its alloys undergo temperature-driven martensitic phase transformation leading to the development of complex microstructures at mesoscale. Optimizing the mechanical properties of these materials requires an understanding of the correlations between the processing parameters and the mechanisms involved in the microstructure formation and evolution. In this work, we study the temperature-induced phase transition from BCC to HCP in pure titanium by atomistic modeling and investigate the influence of local stress conditions on the final martensite morphology. We simulate the transition under different stress conditions and carry a detailed analysis of the microstructural evolution during transition using a deformation gradient map that characterizes the local lattice distortion. \textcolor{black}{The analysis of final martensite morphologies shows how mechanical constraints influence the number of selected variants and the number/type of defects in the final microstructure. We give insight on the origin and structure of different interfaces experimentally observed, such as inter-variant boundaries and antiphase defects. In particular, we show how antiphase defects originate from the two-fold degeneracy shuffling displacement arriving during the transition and how the triple junction formation drives the texture evolution when local stresses prevent a free shape change of the matrix surrounding the growing martensite nuclei.}\end{abstract}


\end{frontmatter}

\linenumbers

\section{Introduction}\label{sec:intro}
Martensitic transformation (MT) is a particular sub-class of solid-to-solid structural phase transformations observed in many metals and alloys~\cite{Bhattacharya2003-lk}. In most general terms, the MT is a diffusionless displacive first-order phase transition. It involves a shear-dominated change of shape in the underlying crystal lattice on alteration of the external conditions, i.e., temperature and/or pressure or stress. Often, a complex microstructure governed by the symmetry of the different phases develops \cite{Khachaturyan1967-zz,Schryvers2002-sm,Delville2009-ed,Finel2010-rd,Salman2012-zg,Salmant,Vattre2016-ql,Salman2019-cg}, giving rise to exceptional mechanical properties such as shape-memory effect \cite{Hildebrand2008-vy}, superelasticity \cite{Tanaka1488} and high-strength \cite{Saito2003-iv,Zhang2017-mr}. In particular, materials of strong interest for the nuclear \cite{lemaignan2006zirconium}, aeronautic \cite{benmhenni2013micromechanical} and bio-medical industries \cite{leyens2003titanium,brunette2012titanium}, such as titanium, zirconium and their alloys, undergo martensitic phase transformation. The present paper specifically deals with martensitic phase transition in pure titanium. However, the observations and conclusions of our work are potentially valid, at least in general terms, also for elements such as zirconium and other alloys undergoing the BCC$\rightarrow$HCP transition.\\
Titanium exists as Hexagonal Close Packed (HCP) phase ($\alpha$-phase) at room temperature and atmospheric pressure. On raising the pressure, while keeping the temperature constant, it transforms to a hexagonal phase ($\omega$ phase) around 2 GPa pressure \cite{banerjee2010phase}. When the temperature is raised at atmospheric pressure conditions, the HCP structure transforms to a Body-Centered Cubic (BCC) structure ($\beta$ phase) at 1155 K,  stabilized by lattice vibrations \cite{heiming1991phonon}. It is clear that, during any conventional transformation route (e.g., metal forming \cite{cabus2014influence}) or advanced elaboration processes (e.g., additive manufacturing \cite{chastand2018comparative,thijs2013strong,thijs2010study}), the transition between the three possible phases can generally occur multiple times, combined with other metallurgical and mechanical phenomena such as plasticity \cite{Conti2004-sv,Bhattacharya2004-es,otsuka2018fft} or recrystallization \cite{Jedrychowski2015-jq}. This fact induces significant modifications of the material microstructure and, as a result, of its mechanical properties. Optimizing these properties requires a clear understanding of the correlations between the processing parameters and the mechanisms involved in the microstructure formation and evolution.\\
Despite the large number of numerical and experimental studies, the debate is still open on topics such as the relationship between initial and final phases \cite{cabus2007orientation,humbert2015evaluation,Chen2016-ue}, the variant selection criterion in bulk material and at grain boundaries \cite{gao2014diffuse,wang2003effect,lischewski2011nucleation,Shi2013-sl,furuhara2001variant}, the exact kinetic and sequence of transformation events \cite{srivastava1995evolution,farabi2018five}, the structure of interfaces between different variant domains \cite{banerjee1998substructure,matsuda2008crystallography,otsuka2005physical}. \textcolor{black}{The present work aims to partially fill this gap of knowledge in the case of the temperature-driven BCC$\rightarrow$HCP phase transition in pure titanium and, in particular, performs a computational investigation on martensite microstructural evolution during the transition and its final morphologies growing under different stress conditions. To our knowledge, this is the first atomistic study investigating such processes in pure titanium.\\
We employ atomistic modeling to simulate the transition under different stress conditions. Contrary to microstructural models written at the mesoscale, this modeling technique does not either exclude any local mechanism or preselect a specific kinetic pathway. Thus, it is well suited to gain insight into the details of the transition. In particular, we used a recently developed atomistic approach based on the overdamped Langevin dynamics \cite{baruffit,baruffi2019overdamped}, which we here apply to simulate a displacive solid-state phase transition.\\
As opposed to experiments, which are mostly limited to the observation of final microstructures, modeling allows to follow the entire microstructural evolution during the transition. We analyse the sequence of nucleation events that subsequently lead to long-stage microstructures emergence under different stress conditions. The analysis of martensite morphologies resulting form the transformation confirms the importance of different stress conditions in the variant selection process and, consequently, on the appearance of different defects (e.g., intervariant boundaries and antiphase defects).Two types of interfaces are mainly observed. The first ones are a consequence of the shuffle degeneracy within each orientational variant. As they do not involve any elastic accomodation, these interfaces are essentially wavy. The second ones result from the impingement between different orientational variants and, consequently, display plane morphologies due to long-range elastic relaxation.}
\begin{table*}[htpb]
	\centering
	\begin{tabular}{ccc}
		\hline
		\multicolumn{3}{c}{STRETCH TENSORS $\bm{U}^{(k)}$} \\
		\hline %
		$ $$\bm{U}^{(1)} = \frac{1}{2}
		\begin{pmatrix}
		2\eta_1   & 0    & 0 \\
		0   & {\eta_2 + \eta_3}   & {\eta_3 - \eta_2} \\
		0   & {\eta_3 - \eta_2}   & {\eta_2 + \eta_3}
		\end{pmatrix}$$ $& $ $$\bm{U}^{(2)} = \frac{1}{2}
		\begin{pmatrix}
		2\eta_1   & 0    & 0 \\
		0   & {\eta_2 + \eta_3}   & {\eta_2 - \eta_3} \\
		0   & {\eta_2 - \eta_3}   & {\eta_2 + \eta_3}
		\end{pmatrix}$$ $   &
		$ $$\bm{U}^{(3)} = \frac{1}{2}
		\begin{pmatrix}
		\eta_2 + \eta_3   & 0    & \eta_3 - \eta_2 \\
		0   &  2 \eta_1 & 0 \\
		\eta_3 - \eta_2  & 0   & {\eta_2 + \eta_3}
		\end{pmatrix}$$ $ \\
		$ $$\bm{U}^{(4)} = \frac{1}{2}
		\begin{pmatrix}
		\eta_2 + \eta_3   & 0    & \eta_2 - \eta_3 \\
		0   &  2 \eta_1 & 0 \\
		\eta_2 - \eta_3  & 0   & {\eta_2 + \eta_3}
		\end{pmatrix}$$ $ &
		$ $$\bm{U}^{(5)} = \frac{1}{2}
		\begin{pmatrix}
		\eta_2 + \eta_3   & -\eta_2 + \eta_3 & 0 \\
		-\eta_2 + \eta_3  &   \eta_2 + \eta_3 & 0 \\
		0 & 0    &2  \eta_1
		\end{pmatrix}$$ $ &
		$ $$\bm{U}^{(6)} = \frac{1}{2}
		\begin{pmatrix}
		\eta_2 + \eta_3    & \eta_2 - \eta_3 & 0\\
		\eta_2 - \eta_3   &   \eta_2 + \eta_3 & 0 \\
		0 & 0    & 2 \eta_1
		\end{pmatrix}$$ $ \\
		\hline
	\end{tabular}
	\caption{The six transformation stretch tensors associated with the BCC$\rightarrow$HCP displacive transformation, written in orthonormal basis aligned with the BCC lattice cubic directions. Coefficients $\eta_1$, $\eta_2$ and $\eta_3$ are related to the BCC and HCP lattice parameters by $\eta_1 = \frac{a}{a_0}$, $\eta_2 = \sqrt{\frac{3}{2}}\frac{a}{a_0}$, $\eta_3 = \frac{c}{\sqrt{2} a_0}$, where $a_0$ and $(a,c)$ are the lattice parameters of the BCC and HCP lattices.}
	\label{tab:U}
\end{table*}
\section{Methods}\label{sec:met}
\subsection{Modeling approach}\label{sec:model}
We employed a recently introduced modeling approach describing the evolution of particle positions with an overdamped stochastic dynamics \cite{baruffit,baruffi2019overdamped} to model the BCC$\rightarrow$HCP transition under different stress conditions. With this method, particle positions are treated as stochastic variables that follow a first-order in time dynamics that do not explicitly incorporate high-frequency vibrations of the crystalline grid (phonons), which limits the time scale of classical Molecular Dynamics to a few nanoseconds  \cite{paul1993molecular}. The chaotic nature of the Newtonian dynamics, which in the long time drives the system to a stochastic equilibrium state, is recovered in the first-order in time dynamics through the use of an additive noise term, carefully chosen to guarantee that the system converges to the correct thermodynamical state in the long-time limit. \textcolor{black}{While this approach has been widely used in studying the dynamics of soft matter systems as well as in biomolecular simulations \cite{ando2003multiple,manghi2006hydrodynamic,ma2016generalized}, it has, to our knowledge, never been employed for the simulation of crystalline materials.} In this section, we report the main equations used in the model. The reader may refer to Refs.~\cite{baruffit,baruffi2019overdamped} for more details on its analytical derivation.\\
In the proposed approach, the configurational space is restricted to the coordinates $x_i^n$, where the upper index $n=1,...,N$  refers to a particle and the lower index $i=1,2,3$ to a cartesian coordinate. Correspondingly, the dynamics involves only the first derivatives of $x_i^n$ and reads as
\begin{linenomath*}
 \begin{equation}
 \frac{dx_i^n}{dt} = - \nu^{-1} {\frac {\partial \Phi}{\partial x_i^n}} +  B \eta_i^n(t),
 \label{eq:lg}
 \end{equation}
 \end{linenomath*}
 where $\Phi(\{ x_i^n\})$ is the potential energy between particles, $\nu$ a viscosity coefficient and $B$ the amplitude of a white Gaussian noise $ \eta_i^n(t)$ such that $\langle\eta_i^n(t)\rangle=0$, $\langle\eta_i^n(t)\eta_j^m(t')\rangle=\delta_{nm}\delta_{ij}\delta(t-t')$. $\delta_{ij}$ and $\delta_{ij}$ are Kronecker symbols and  $\delta(t-t')$ stands for the Dirac-delta distribution. Equations (\ref{eq:lg}) represent a first-order in time stochastic dynamics, also known as overdamped Langevin Dynamics~\cite{Nelson1967-ee}. We hypothesize that the coefficients $\nu$ and \textit{B} are  independent from particle positions and related by a fluctuation-dissipation relation, i.e., $B=\sqrt{2k_BT\nu^{-1}}$. This guarantees that, in the long-time limit $t \rightarrow \infty$, the distribution probability $P (\{x_i^{n}\})$ generated by Eqs.~(\ref{eq:lg}) converges to a steady state characterized by the Boltzmann distribution
\begin{linenomath*}
\begin{equation}
	t \rightarrow \infty: \ \ \ \ \ \  P(\{x_i^{n}\}) \rightarrow A \exp \left( - {\frac{\Phi (\{x_i^{n}\})}{k_B T}}\right).
\end{equation}
\end{linenomath*}
Supplemented by the constraint that the particles stay within a pre-defined simulation box, the dynamics represented by the set of equations (\ref{eq:lg}) is valid in the $(NVT)$ thermodynamical ensemble, i.e., the number of particle $N$, the volume $V$ and the temperature $T$ are fixed. To deal with applied stress conditions, we extended the model to the $(N\bm{P}T)$ ensemble, where $\bm P$ stands for the first Piola-Kirchhoff tensor. We present now briefly the stochastic dynamics required for this $(N\bm{P}T)$ ensemble.\\
First, we incorporate nine additional degrees of freedom into the model, which are the components of the deformation gradient $\bm{F}$ describing the change in the shape of the simulation box. Next, in order to couple $\bm{F}$ to the degrees of freedom associated to the atomic positions, we introduce scaled coordinates $\{\tilde{x}_i^n\}$ related to the actual coordinates by
\begin{linenomath*}
\begin{equation}
\tilde{x}_i^n = \left( \bm{H}^{-1}\right)_{ij} x^n_j, \ \ \ \ \ \ \ i=1,2,3
\end{equation}
\end{linenomath*}
where the matrix $\bm{H}$ is defined by $\bm{H} = \bm{F} \bm{L}^0$, where $\bm{L}^0$ is a diagonal matrix containing the length of the orthogonal vectors $\bm{L}^0_1$, $\bm{L}^0_2$ and $\bm{L}^0_3$ that define the initial simulation box. The extended overdamped Langevin dynamics reads as
\begin{linenomath*}
\begin{align}
\nonumber
\frac{d \tilde{x}_i^n}{dt} & = - \nu^{-1}{\frac {\partial \tilde{H}}{\partial \tilde{x}_i^n}} + \sqrt{2 k_B T \nu^{-1}}\, \eta_i^n(t) \ \ \ i=1,2,3 \ ; \ n=1,...\text{N} \ , \\
\frac{d F_{ij}}{dt} & = -  \gamma^{-1} {\frac {\partial \tilde{H}}{\partial F_{ij}}} + \sqrt{2 k_B T \gamma^{-1}} \, \xi_{ij}(t) \ \ \ i,j = 1, ..., 3 \ ,
\label{eq:lg_ext}
\end{align}
\end{linenomath*}
where $\xi_{ij}(t)$ is a white Gaussian noise with $\langle \xi_{ij}(t) \rangle=0$ and $\langle \xi_{ij}(t) \xi_{kl}(t') \rangle=\delta_{ik}\delta_{jl}\delta(t-t')$, $\gamma$ a viscosity associated with the new degrees of freedom $F_{ij}$ and $\tilde{H}(\{ \tilde{x}_i^n \},\bm{F})$ the Hamiltonian for the extended set of DOF. The extended Hamiltonian should of course be such that Eqs.~(\ref{eq:lg_ext}) converge in the long time limit towards the thermodynamical equilibrium of the ($N\bm{P}T$) ensemble. As shown in \cite{baruffi2019overdamped}, this leads to:
\begin{linenomath*}
\begin{equation}
\tilde{H}(\{\tilde{x}_i^n\}, \bm{F}) = \Phi(\{F_{ij}L_j^0 \tilde{x}_j^n\}) + V_0 P_{ij} F_{ij} - N k_B T \ln \left( V_0 \det \bm{F}\right),
\label{eq:log}
\end{equation}
\end{linenomath*}
where $V_0$ is the volume of the initial simulation box. \textcolor{black}{We note that this extended Hamiltonian is equal to the usual enthalpy\footnote{\textcolor{black}{Note that our sign convention is such that $P_{ij}= \delta_{ij} P$ where $P > 0$ corresponds to a system under hydrostatic pressure.}}
 $(\Phi + V_0 \bm{P} \bm{F})$ supplemented by an extra logarithmic term, which ensures that the Langevin dynamics converges towards the correct thermodynamic equilibrium \cite{baruffi2019overdamped}.
 In principle, this logarithmic term should include a normalization volume unit $V_B$ that would make the argument of the logarithmic term dimensionless but, as long as we are considering kinetics and the associated driving forces, this normalizing term plays no role\footnote{\textcolor{black}{In principle, the logarithmic term in Eq.~\eqref{eq:log} should be written as $\log \left( V_0/V_B \det \bm{F}\right)$ where the quantity $V_B$, which has the dimension of a volume, is given by $V_B = \Lambda^3$ where $\Lambda$ is the Broglie wavelength. This term is reminiscent of the quantum and, therefore, discrete nature of the Hamiltonian and should be incorporated if we want to ensure a proper normalization of the entropies and energies. However, as it enters only through a constant term that disappears through derivation, this normalizing term does not enter the driving forces and plays no role within the dynamics. This is the reason why we do not include it in Eq.~\eqref{eq:log}, even though it would make the argument of the logarithmic term dimensionless.}} and, therefore, is not included in Eq.~\eqref{eq:log}. Altogether, the use of the extended Hamiltonian within Eq.~\eqref{eq:lg_ext} guarantees that the dynamics converges to the correct equilibrium state associated with the $(N\bm{P}T)$ ensemble in which an externally applied first Piola-Kirchhoff stress controls the system. We stress that the appearance of the first Piola-Kirchhoff stress is simply a consequence of the fact that this is the stress measure conjugated to the deformation gradient $\bm{F}$. The deformation gradient itself emerges because, within an atomic-scale approach that naturally uses atomic coordinates that refer to a fixed reference state, it is most convenient to use a Lagrangian description within which the degrees of freedom of the fluctuating simulation box are simply the entries of the deformation gradient $\bm{F}$ (see for example \cite{miller2016molecular}). Finally, we mention that, even though it is not required for the implementation and integration of the kinetic equations, we could define an instantaneous internal first Piola-Kirchhoff stress whose statistical average adopts a virial form which, at equilibrium, is automatically equal to the externally applied Piola-Kirchhoff stress (see Appendix A for details).}\\
\textcolor{black}{We conclude this paragraph with a general remark on the proposed overdamped Langevin dynamics. Such dynamics, also known as Brownian dynamics, have already been used for the modeling of a number of spatio-temporal processes, in which heavy particles (such as biomolecules) interact with a bath of light particles (e.g. solvent molecules). In the limit of vanishing ratio $m/M$, where $m$ and $M$ are the masses of the light and heavy particles, it can be argued that the heavy particles follow a Markovian dynamics, which permits to exclude from the simulation the light particles, which are of no direct interest (see for example \cite{erban2014molecular}). The present situation is different, as the atomic species that constitute our materials cannot be separated into subclasses with different masses. However, our aim is to show that, because of the randomness generated by the chaotic character of the phase space trajectories, the initial deterministic dynamics, which is of second order in time, can be replaced by a first order in time stochastic dynamics. Of course, an exact formulation of the overdamped Langevin equations should proceed through an explicit coarse-graining procedure over the initial Newtonian dynamics, which requires the identification of a characteristic time over which time averages can be performed while preserving the characteristic times associated to the dynamics processes we are interested in. This coarse-graining procedure would naturally lead to a coarse grained potential $\Phi_{cg}\left (\{x^n_i\} \right )$ that will differ from the original potential $\Phi \left(\{x^n_i\}\right)$, as phonons will be adiabatically embedded into $\Phi_{cg}\left (\{x^n_i\} \right )$, together with explicit expressions for the noise terms and for the viscosity coefficient $\nu$, whose knowledge is needed to access to the time scale of our first order in time dynamics. Because of the intrinsically anharmonic character of the initial potential, we anticipate that the coarse-grained potential will be significantly softer than the original one, allowing to use large time step when the dynamics is numerically discretised. In this paper, we propose a simplification that consists in replacing the coarse-grained potential by the original one. We therefore cannot associate a real time scale to our simulations. However, we show below that our first-order in time dynamics does reproduce the real nature of the dynamics in terms of reaction pathway and observed microstructures. This asserts that a first-order in time out-of-lattice dynamics can be safely used to simulate spatio-temporal processes even though the underlying atomic species cannot be separated into subclasses associated to different characteristic times.}
\subsection{Numerical implementation}
\textcolor{black}{The model described in section \ref{sec:model} has been implemented in a Fortran code of our own. We referred to the work of Goedecker \cite{goedecker2002optimization} to optimize interatomic forces computation by parallel computing. To time integrate Eqs.~(\ref{eq:lg_ext}), we used an explicit predictor-corrector method \cite{Burrage2007-gj,baruffi2019overdamped}. 
Dimensionless equations are obtained by introducing an energy unit $E_0$ and a time unit $t_0$ = $\nu / E_0$, where $\nu$ is the viscosity term that enters into the dynamics of the scaled atomic coordinates. When needed, we display our results with reference to the dimensionless time $\tau = t / t_0$. For simulation in the $(NVT)$ ensemble, the time integration is performed using an explicit predictor-corrector method with a dimensionless time step $\Delta \tau = 10^{-5}$. For the  $(N\bm{P}T)$ ensemble, the simulation box viscosity $\gamma$ is set equal to 0.0855 $\nu$ and the dimensionless time step is $\Delta \tau = 10^{-7}$. More details on the numerical integration of Eqs.~(\ref{eq:lg_ext}) are given in Appendix B.}
%
%
\subsection{Simulation setup and interatomic potential}
To clarify the influence of different external conditions on martensite microstructure, we simulate the BCC$\rightarrow$HCP transition in both thermodynamic ensembles $(NVT)$ and $(N\bm{P}T)$.
\textcolor{black}{We use periodic boundary conditions. In the (NPT) ensemble, we consider stress-free conditions by setting to zero the first Piola-Kirchhoff stress, allowing the material to change its macroscopic shape.}
Although real conditions experienced by a region in bulk material would be an intermediate case between these two conditions, the two extreme scenarios are useful for a global understanding of the influence of local mechanical constraints preventing a free change in shape and/or volume of the matrix around a martensite nucleus.

In the simulations, we first equilibrate BCC titanium at $1400$ K and then quench it at $700$ K. We perform the quenching by an instantaneous rescaling of the temperature parameter that fixes the noise amplitudes in Eqs. \ref{eq:lg_ext}. The simulation box size is set equal to $36 \times 36 \times 36$ $a_0^3$, where $a_0$ is the BCC equilibrium lattice constant at 1400 K ($a_0=3.417 \; \angstrom$). The total number of atoms is 93312.

\textcolor{black}{But first, to perform Molecular Dynamics or Langevin atomic simulations, a relevant interatomic potential is needed. We have considered two empirical atomic potentials for titanium from the literature that could be relevant because they were in particular developed to study the BCC$\rightarrow$HCP transition.
The first potential, of the EAM type, proposed by Mendelev et al.~\cite{mendelev2016development} and referred to as Ti-1 EAM, was fitted to reproduce the HCP stacking fault energy, the BCC-HCP transformation temperature ($T \sim 1150$K) and the melting temperature.
The second potential, of the MEAM type, proposed by Henning et al.~ \cite{hennig2008classical} describes the structure and energetics of $\alpha$, $\beta$ and $\omega$ phases in Ti. Optimization of the parameters is performed using a database of density-functional calculations and  yields an accurate potential as verified by comparison to experimental and density-functional data for phonons, surface and stacking fault energies, and energy barriers for homogeneous martensitic transformations. In addition, the elastic constants, phonon frequencies, surface energies, and defect formation energies closely match density-functional results even when these were not included in the fitting procedure. The authors have also verified using Molecular Dynamics that the equilibrium phase diagram is in close agreement with experimental measurements.}

We tested these EAM and MEAM potentials by performing preliminary simulations with classical Molecular Dynamics using the simulator LAMMPS~\cite{Plimpton1995-wa}. We obtained the following results: when the MEAM-type potential is used, we were able to observe a stable BCC phase transforming into HCP upon quenching, coherently with previous Molecular Dynamics simulations proving its ability to reproduce the whole temperature-pressure phase diagram of titanium \cite{hennig2008classical}. On the other hand, when using the EAM potential we did not observe phase transition after cooling, although we were able to get a stable BCC structure at high temperatures. We increased the simulation duration up to 1 nanosecond and we tested different simulation box sizes. However, the transition did not occur in any of the two thermodynamic ensembles. The possible reasons for that could be: i) the simpler functional form of the EAM potential compared to the MEAM. The lack of any angular dependency in the embedding term describing the electron density makes the EAM much cheaper than the MEAM from a computational point of view but it impacts its ability to model materials with strong bond directionality metals with partially full-\textit{d} shell, ii) the presence of a high energy barrier for the nucleation of the HCP phase.

Based on these results, we finally decided to use the MEAM potential to perform the overdamped Langevin simulations and implemented it in a parallel code by following a previous work on many-body force field implementation~\cite{goedecker2002optimization}. 

\subsection{Variant and phase identification}
To characterize the microstructure formation and evolution, we need to identify the different crystal structures and, especially, the different variants.\\
To identify the different HCP variants, we use a deformation gradient map representing the local lattice distortion. Indeed, when a material undergoes a martensitic transformation, several energetically equivalent variants differing in their relative crystallographic orientation appear \cite{pitteri2002continuum}. Each of these variants is associated with a stretch tensor $\bm{U}$ that can be easily identified once the local deformation gradient $\bm{F}$ is known: we just need to use the polar decomposition $\bm{F} = \bm{Q} \bm{U}$ (where $\bm{Q}$ is a rotation and $\bm{U}= \bm{U}^T$ is positive-definite), which is unique. However, due to the infinite degeneracy of the lattice groups of the parent and product phases, the identification of the deformation gradient $\bm{F}$ is not unique. Therefore, it is common to use a lattice correspondence between the parent and product phases to represent the actual lattice sites' displacements \cite{banerjee2010phase}. Before explaining the procedure used to identify $\bm{F}$, we first recall that a simple homogeneous deformation gradient cannot fully describe the martensitic transformation from BCC to HCP: the BCC lattice is a Bravais lattice, the HCP is not. Therefore, the transformation strain that we want to identify must be supplemented by atomic displacements applied on a sublattice of the deformed lattice. In the present situation, this shuffling consists in translating every second basal plane of the hexagonal lattice obtained after the homogeneous deformation gradient.

We now turn to the procedure used to identify the local deformation gradient. This identification requires a lattice correspondence between the parent and product phases. Two lattice correspondences, given in terms of orientation relationships, have been proposed. \textcolor{black}{The mechanism given by Burgers \cite{burgers1934process} states that the following crystallographic planes and directions are parallel:}
\begin{linenomath*}
\begin{equation}
(110)_{bcc} \parallel (0001)_{hcp};[\bar{1}11]_{bcc}  \parallel [\bar{2}110]_{hcp},
\label{eq:burgers_text}
\end{equation}
\end{linenomath*}
whereas the mechanism given by Mao \cite{mao1967effect} states the following correspondence:
\begin{linenomath*}
\begin{equation}
(110)_{bcc}  \parallel (0001)_{hcp};[00\bar{1}]_{bcc} \parallel [11\bar{2}0]_{hcp}.
\label{eq:mao_text}
\end{equation}
\end{linenomath*}
The two mechanisms differ only in that the Burgers mechanism requires a rotation of $\pm 5.26^\circ$ around the $[0001]$ HCP axis in order to obtain the proposed direction correspondence \cite{wang1998iron}. Consequently, whereas the Mao relationship generates only 6 HCP orientation variants, the Burgers mechanism generates 12 HCP lattices. However, as they differ only by rotations, the two mechanisms are associated with exactly the same six stretch tensors $\bm{U}^{(k)}, \ \ k=1, \ldots, 6$. These tensors are listed in Tab.~\ref{tab:U}. We define a procedure meant to identify these local stretch tensors $\bm{U}_n^{(k)}$. For each of the six $(110)$ BCC planes transforming in the final $(0001)$ HCP plane, we define a specific set of neighboring sites, see Appendix C. Then, for each atom \textit{n}, we identify six local deformation gradients $\bm{F}_n^{(k)}$, $k=1, \ldots, 6$, that minimize the following local descriptors:
\begin{linenomath*}
\begin{equation}
k=1, \ldots, 6: \ \  D_n^{{(k)}^2} = \sum_{m \in \Omega_n} \lVert \Delta \mathbf{r}_{nm}(t^*) - \bm{F}_n^{(k)}\Delta, \mathbf{r}_{nm}(0)\lVert^2
\label{eq:Dsq}
\end{equation}
\end{linenomath*}
where $\Omega_n$ is the neighborhood set that is associated with a given $(110)$ plane. The local deformation gradient $\bm{F}_n$ is defined as the one that, among the six tensors $\bm{F}_n^{(k)}$, leads to the smallest $D_n^{(k)}$. Finally, polar decomposition leads to the local stretch tensor $\bm{U}_n$ and to a local stretch deformation map. The non-affine displacement $D_n^{(k)^2}$ quantifies the degree at which an affine transformation can describe the local change in the lattice. In the following analysis, we set a threshold $D^{2}_{lim}=6.5$ \AA\textsuperscript{2} above which the calculated $\bm{F}_n$ is considered not meaningful and exclude from post-processing atoms with $D_n^{(k)^2} > D^{2}_{lim}$.\\
To monitor the evolution of the  phase fraction of  each phase without distinguishing variants, we use the Polyhedral Template Match analysis (PTM)  \cite{larsen2016robust} implemented in the software OVITO \cite{stukowski2009visualization}. This method classifies crystal structures according to the topology of the local atomic environment. It provides a flexible tool for structural identification even in the presence of strong thermal fluctuations when other methods relying on interatomic distances (e.g., Common Neighbor Analysis \cite{honeycutt1987molecular}) are less robust. In our analysis, the cut-off for the Root-Mean-Square-Deviation (RMSD) between the local atomic structure and the ideal structural template has been set equal 0.14 \AA.
\section{Results}\label{sec:ris}
\begin{figure}[t]
	\centering
	\includegraphics[width=0.4\textwidth]{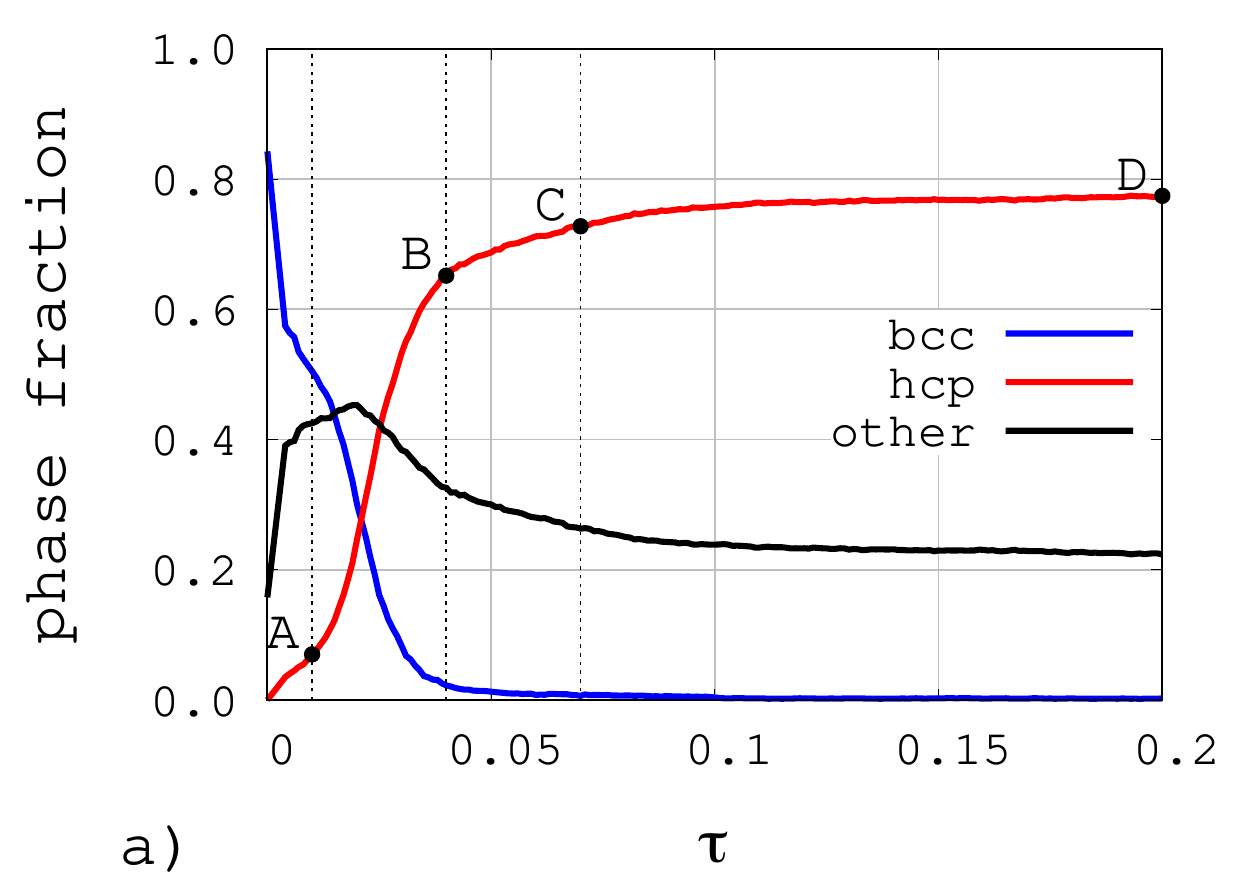}
	\includegraphics[width=0.4\textwidth]{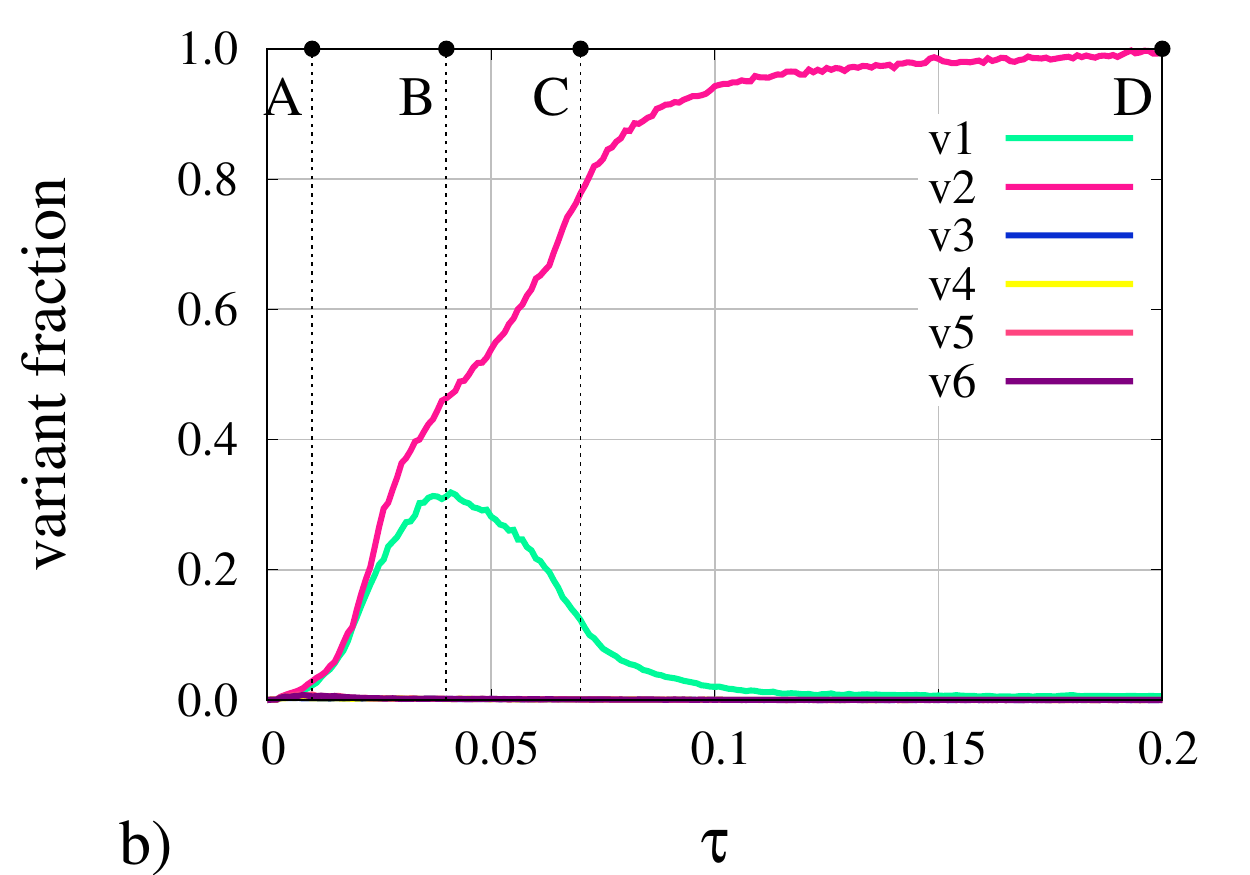}
	\caption{Evolution of BCC and HCP volume fractions (a) and variant fractions (b) in the ($N\bm{P}T$) ensemble. \textcolor{black}{The symbols A,B,C,D indicates the times selected for displaying the microstructures in Fig.\,\ref{fig:npt}.}}
	\label{fig:npt_frac}
\end{figure}
\begin{figure}[h]
	\centering
	\includegraphics[width=0.23\textwidth]{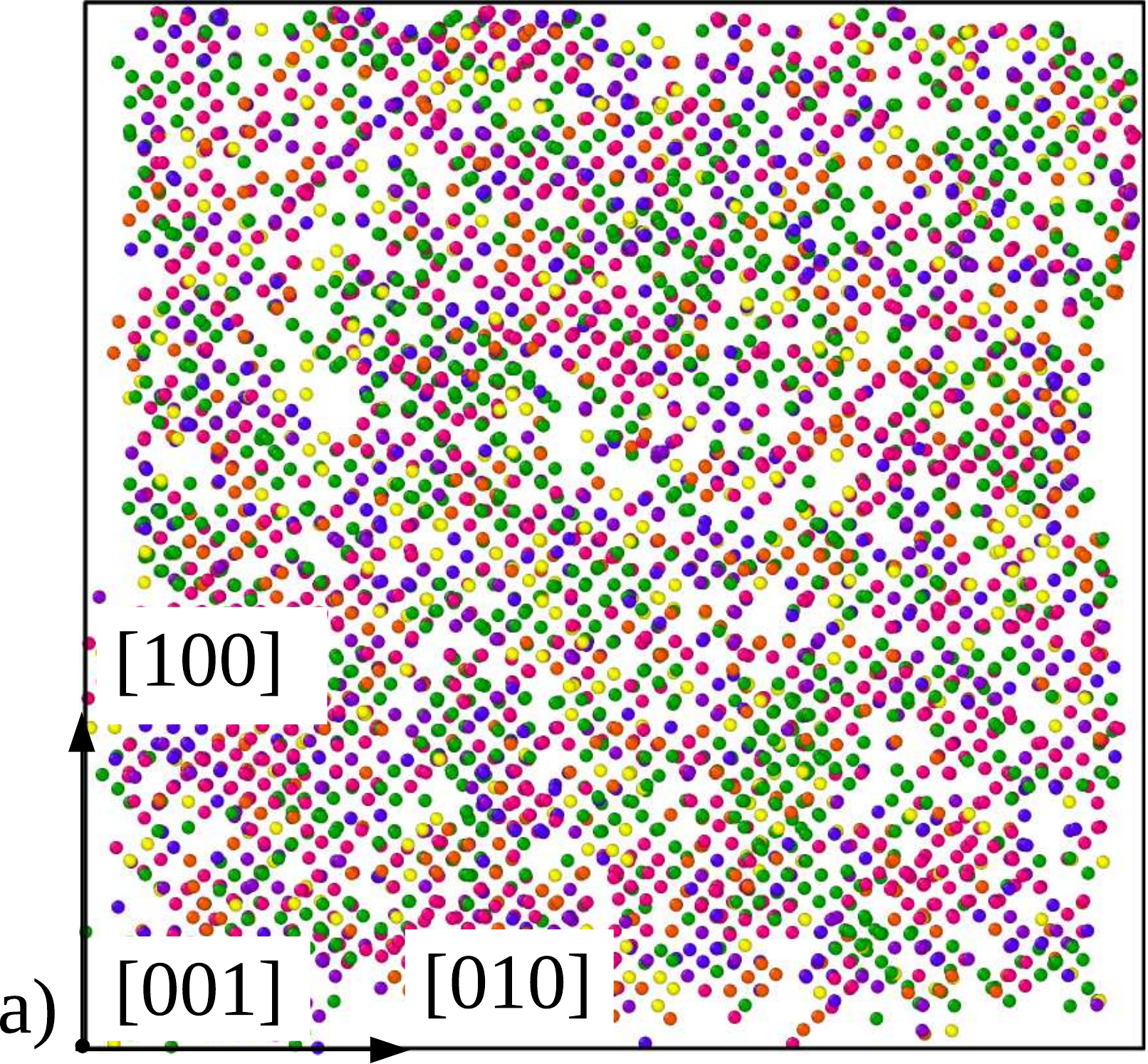}
	\includegraphics[width=0.23\textwidth]{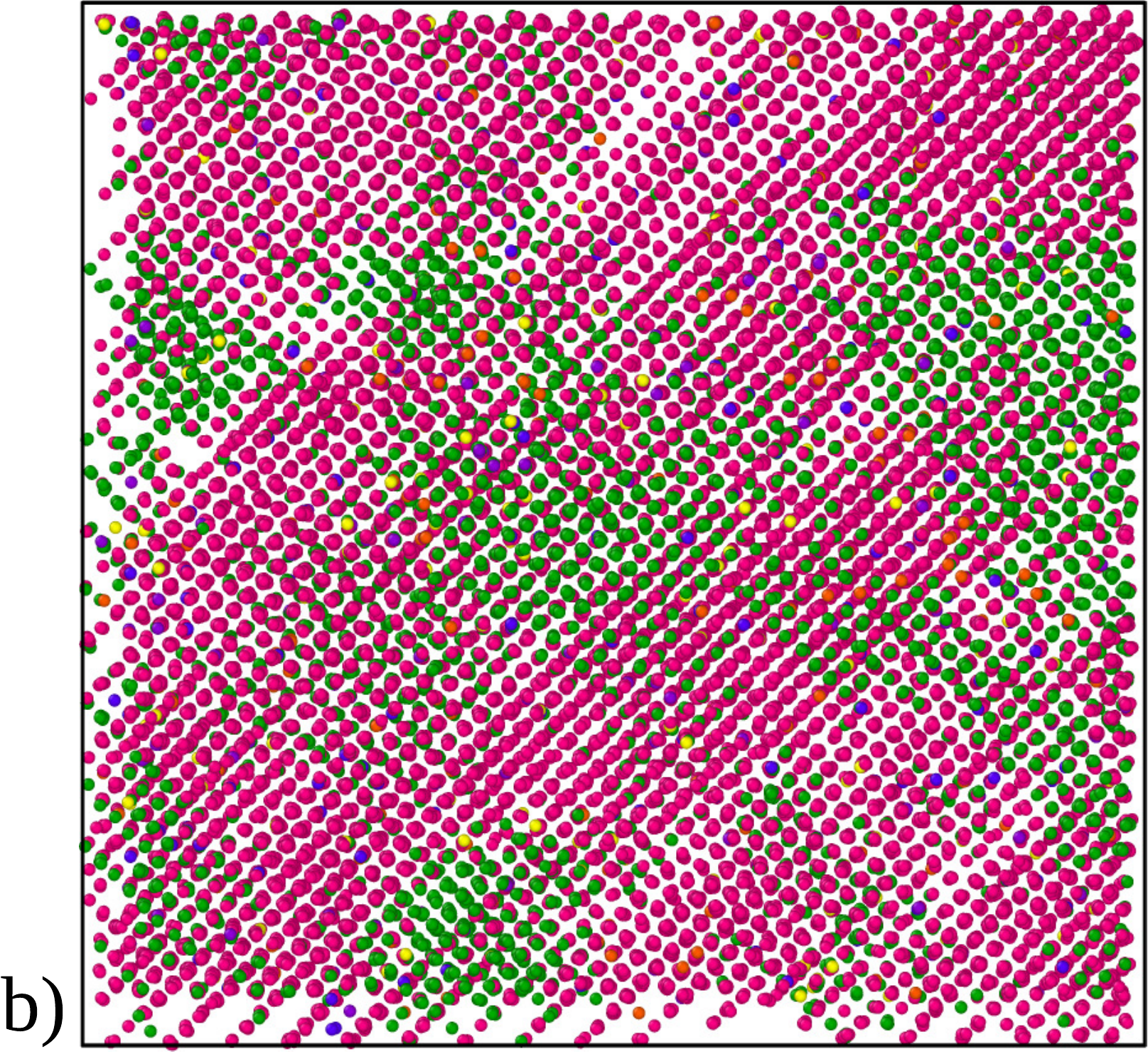}
	\includegraphics[width=0.23\textwidth]{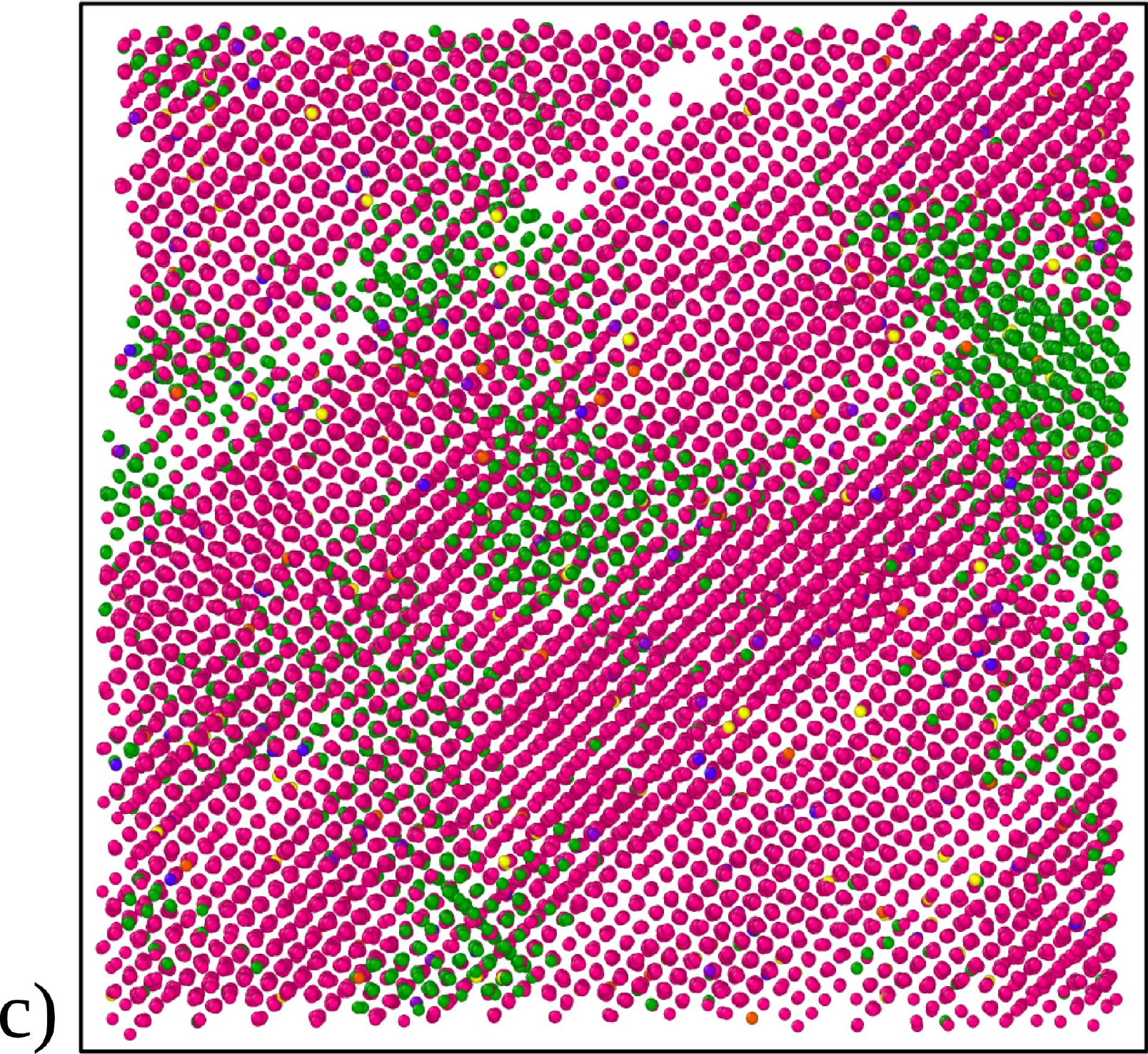}
	\includegraphics[width=0.23\textwidth]{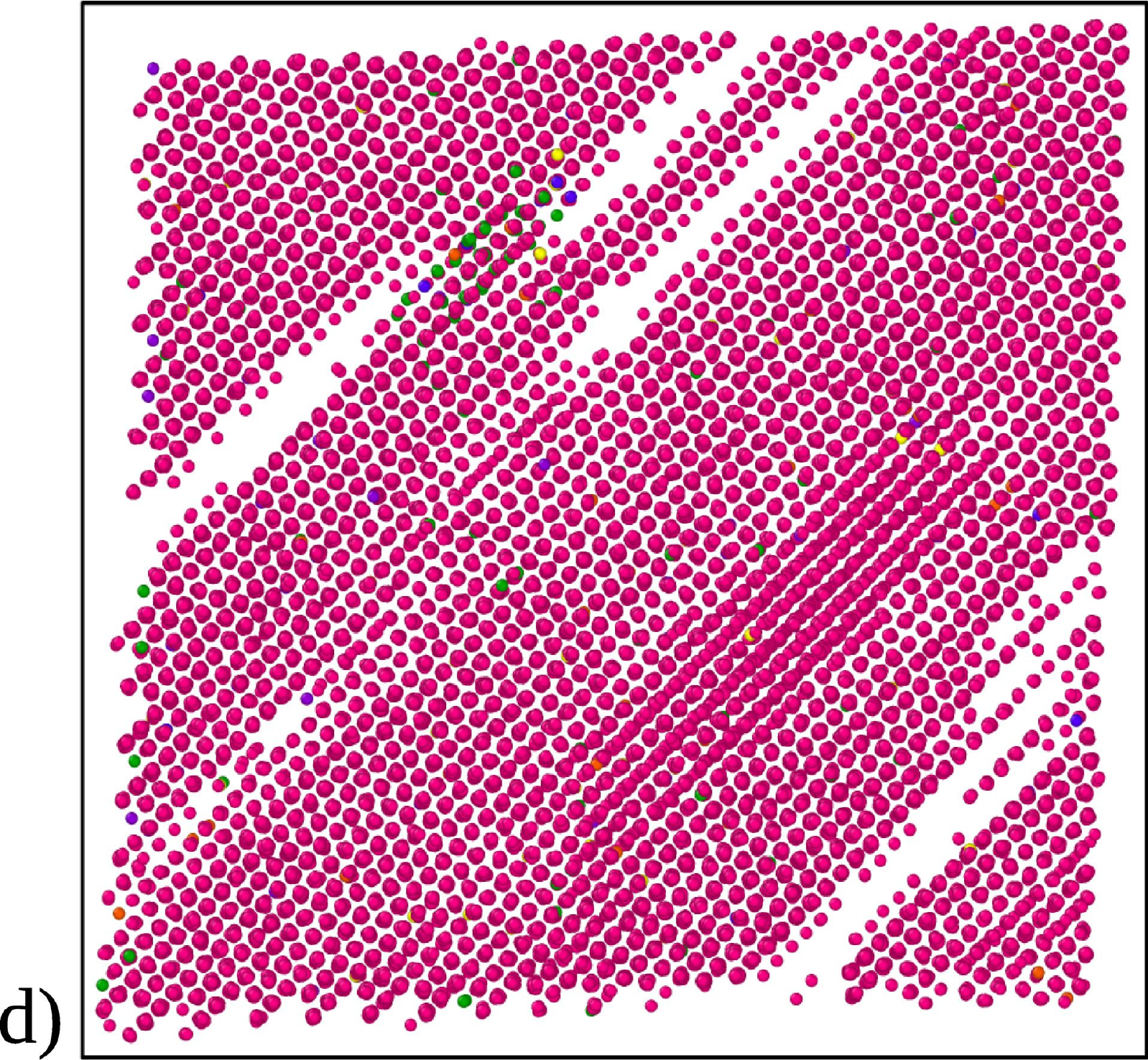}
	\caption{\textcolor{black}{Microstructure evolution in the ($N\bm{P}T$) ensemble. Only atoms within a thin slice of the simulation box and classified as HCP are shown and colored on the basis of the corresponding variant: a) initial nucleation stage, (b)-(c) two variants prevail, (d) final single variant domain. Crystallographic directions refer to the parent BCC phase.The microstructures a,b,c,d correspond to the reduced times 0.01, 0.04, 0.07, 0.2, respectively. These times are indicated in Fig.\ref{fig:npt_frac}
}}
	\label{fig:npt}
\end{figure}
%
%
\subsection{Simulations in the (NPT) ensemble  with stress-free boundary conditions}
\begin{figure} [h]
\centering
{\includegraphics[scale=0.31]{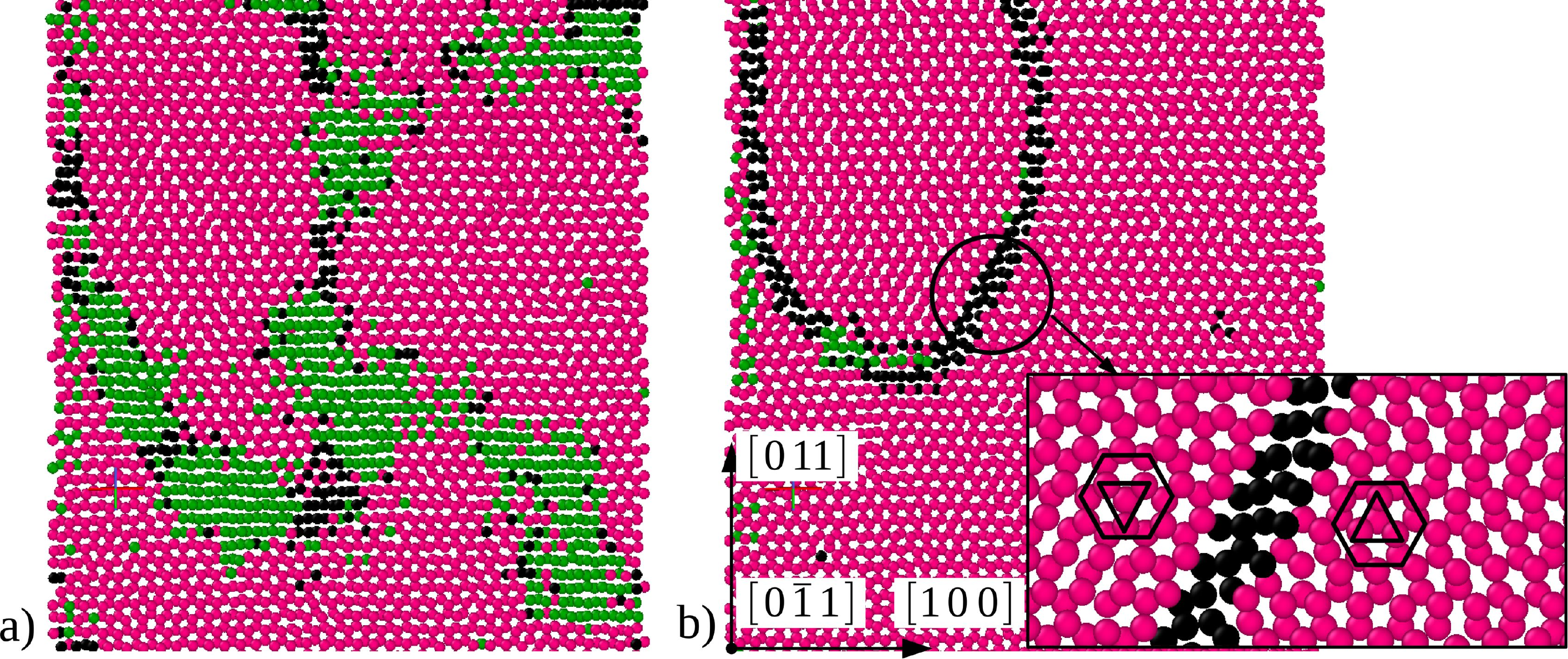}}
\caption{Final state in the $(N\bm{P}T)$ ensemble: atoms with crystallography different from HCP are colored in black: a) anti-variant domains (pink) separated by another variant (green), b) antiphase defect (see the inset) formed at the boundary between anti-variant domains after microstructure coarsened. Crystallographic directions refer to the parent BCC phase.}
\label{fig:antiphase}
\end{figure}

After equilibrating the system in the BCC state at T=1400 K, we drop the temperature down to T=700 K. After a short relaxation, the system transforms into a HCP structure.\\
In Fig.~\ref{fig:npt_frac}a, we report the evolution of the BCC and HCP volume fractions during the transition. Almost no BCC phase is left after the transformation has been completed.  A non-negligible residual fraction of atoms ($\approx$ 0.20) exhibits crystallographic structure different than HCP, which suggests that some defects are generated.\\
In Fig.~\ref{fig:npt_frac}b, we report the evolution of variant fractions as a function of time. At the very beginning of the transition, all the six variants nucleate almost instantaneously. \textcolor{black}{However, very quickly, most of them disappear, giving rise to a microstructure composed of the variants $\bm{U}^{(1)}$ and $\bm{U}^{(5)}$. Afterwards, the structure coarsens further and forms a single variant $\bm{U}^{(5)}$ domain (Fig.~\ref{fig:npt}).}
Snapshots taken during the transition (Fig.~\ref{fig:antiphase}) show that, when the microstructure is coarsening, HCP domains with same \textbf{c} axis orientation but different shuffling directions (referred to as a couple of ``anti-variants'' \cite{gao2014diffuse}) come into contact and generate an antiphase boundary (see the inset Fig.~\ref{fig:antiphase}b). This boundary is equivalent to a stacking fault when the plane between the two domains is parallel to the $\{0001\}$ HCP basal plane. %
%
\subsection{Simulations in the (NVT) ensemble} \label{sec:nvt}
\begin{figure} [h]
\centering
\includegraphics[width=0.4\textwidth]{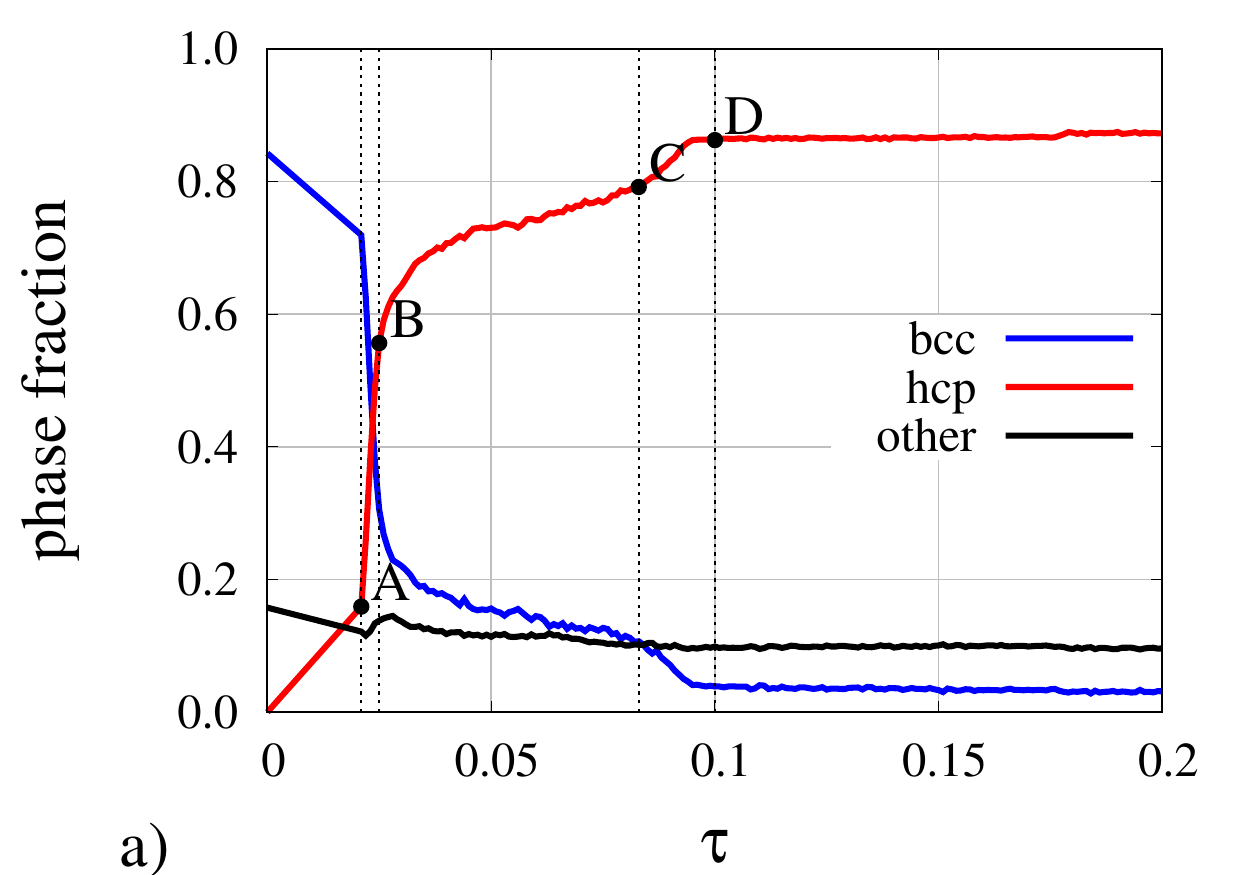}
\includegraphics[width=0.4\textwidth]{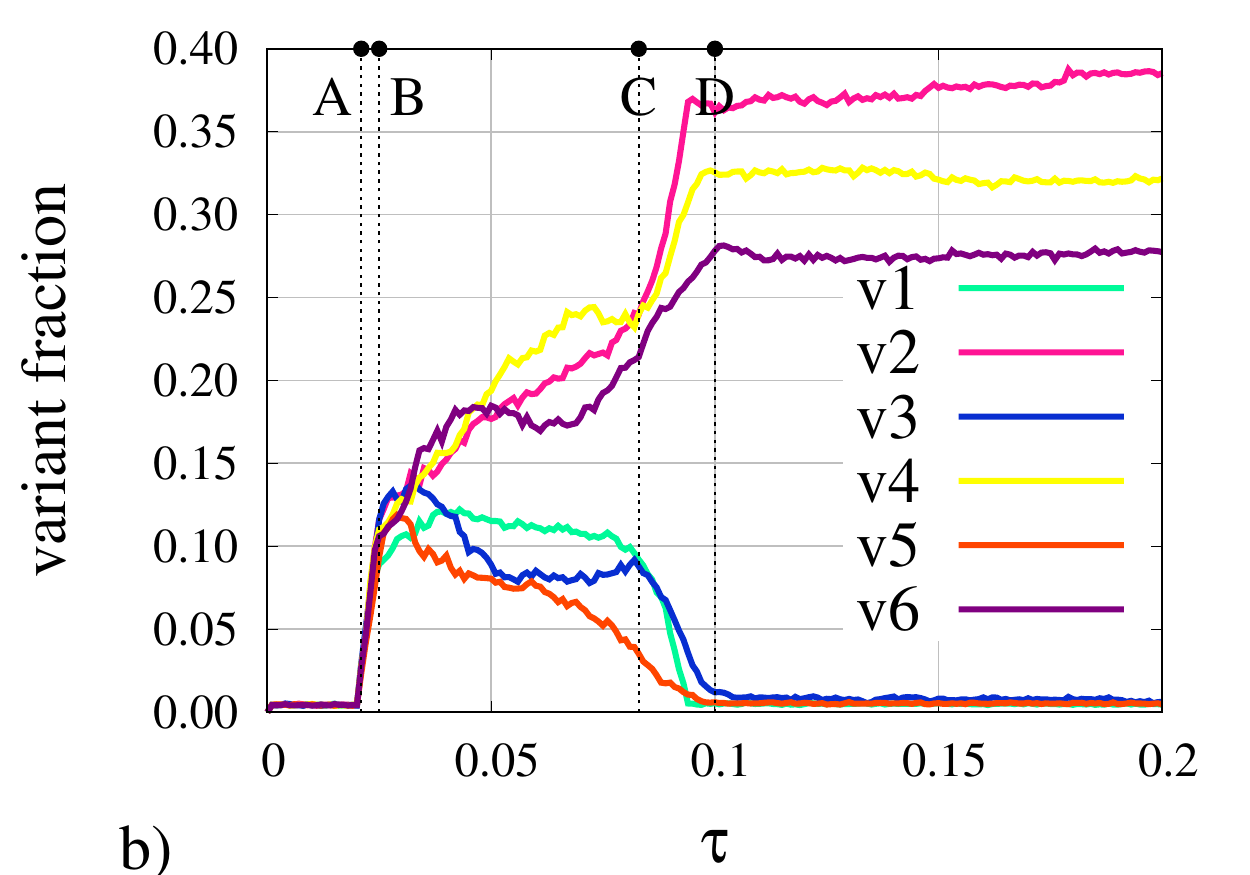}
\caption{Evolution of BCC and HCP volume fractions (a) and variant fractions (b) in the ($NVT$) ensemble.} 
\label{fig:var_sel_4}
\end{figure}

After quenching the system down to $T$=700\,K, the crystal structure transforms into a HCP structure without any remaining BCC domain, similarly to what is observed in the previous $(N\bm{P}T)$ simulation.\\
Fig.~\ref{fig:var_sel_4}a shows the evolution of the BCC and HCP volume fractions  while Fig.~\ref{fig:var_sel_4}b shows the evolution of the six variant volume fractions. The overall transformation proceeds through different stages.\\
We first observe a nucleation stage, that extends up to point A in Fig.~\ref{fig:var_sel_4}a and \ref{fig:var_sel_4}b, during which local HCP fluctuations emerge. We underline that the length scale of these fluctuations is too small to allow the identification of different variants with the procedure outlined in section \ref{sec:met}, thus leading to an apparent incompatibility between Fig.~\ref{fig:var_sel_4}a and Fig.~\ref{fig:var_sel_4}b. Indeed, the procedure used to identify HCP variants relies on neighborhoods that extend beyond the second neighbor shell, whereas the PTM algorithm used to identify the local lattice relies on a neighboring set limited to the first two neighbor shells \cite{larsen2016robust}. Next, HCP nuclei grow rapidly (from point A to point B) and stabilizes to a quasi-stationary stage during which the six HCP variants reach finite volume fractions that are roughly constant (from point B to point C). Then, the system enters a stage during which the volume fraction of three HCP variants increases at the expense of the three others (from C to D). After this growing stage (from point D on),the system stabilizes in a microstructure that consists of only three variants. The selected variants share the $[111]$ BCC direction in the parent phase, i.e., a $\langle 11\bar{2}0 \rangle$ HCP dense direction.\\
\begin{figure} [h]
\centering
{\includegraphics[scale=.32]{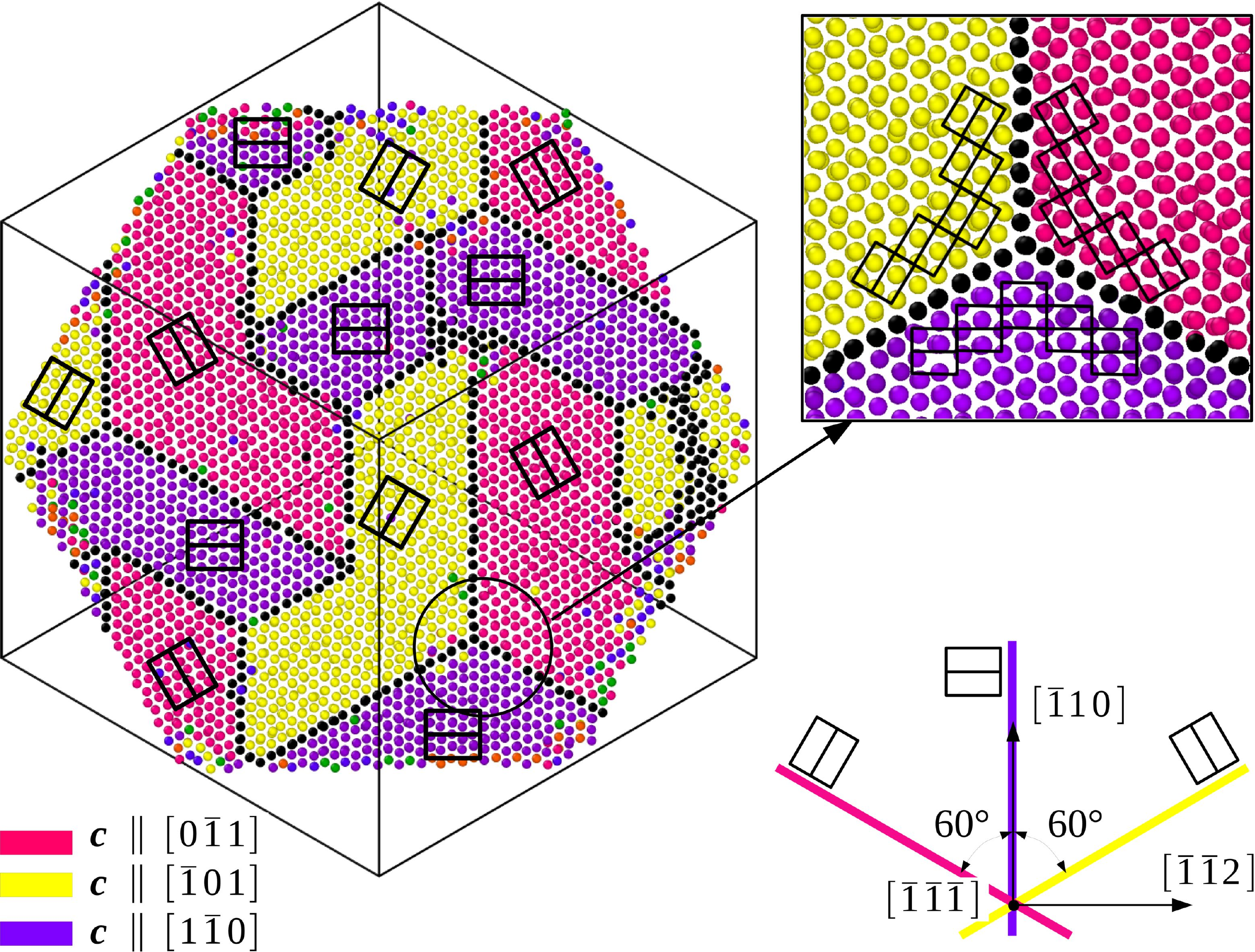}}
\caption{Final microstructure obtained at 700 K in the $(NVT)$ ensemble. The inset shows a triple junction detail. Atoms classified as HCP are colored according to the corresponding variant and atoms with crystallographic symmetry different than HCP in black. Crystallographic directions refer to the parent BCC phase.}
\label{fig:microstru_1}
\end{figure}
Fig.~\ref{fig:microstru_1} shows the final 3-variant microstructure along a plane orthogonal to the common dense HCP direction, within a color map of the \textbf{c} axis orientation and a schematic indication of inter-variant misorientation. \textcolor{black}{The three variants organize around several triple junctions by forming three interfaces which show the structure of the type I $\{10\bar{1}1\}$ twin boundaries (twinning plane $K_1 =\{10\bar{1}1\}$, shear direction $\eta_1 = \langle 1 \bar{2}10\rangle$). The strains associated with the 3 variants forming the 3-plate morphology respect the twinning equations \cite{Bhattacharya2003-lk,bowles1954crystallography}. As later discussed in the Discussion part, this 3-plate geometry is not fully compatible with the formation of three $\{10\bar{1}1\}$ twins so further strain is required for its accomodation.}\\
In Fig.~\ref{fig:nvt}, we show four snapshots of the microstructural evolution during the transition. First, stable nuclei of all the six variants appear (a), and all the different HCP domains develop (b). At this point, two stable triple junctions (indicated by arrows) are already formed and lead to the final 3-plate morphology after the subsequent microstructure coarsening.\\
We repeated the simulation in the $(NVT)$ ensemble several times and changed the random noise term. The time evolution of variant fractions (Fig.~\ref{fig:fig6}) show that in each case the system behaves similarly and, after the nucleation of all the possible variants, progressively selects a triplet. In all the simulations, the selected triplets share a $\langle 111 \rangle$ BCC direction i.e., a $\langle 11\bar{2}0 \rangle$ HCP direction. In terms of microstructure, the selected triplet always organize in the 3-plate morphology. Only in one case, shown in Fig.~\ref{fig:microstru_nvt_2}, the selected variants form two laminates consisting of parallel twins along $\{10\bar{1}1\}$ HCP plane. At the crossing point between the laminates, an FCC domain appears. This FCC domain shares coherent interfaces with the neighboring HCP variants; these sharp interfaces consist in a one-layer thick transition from a $\{111\}$ FCC plane to a HCP basal plane.
\begin{figure}[h]
	\centering
	{\includegraphics[scale=0.28]{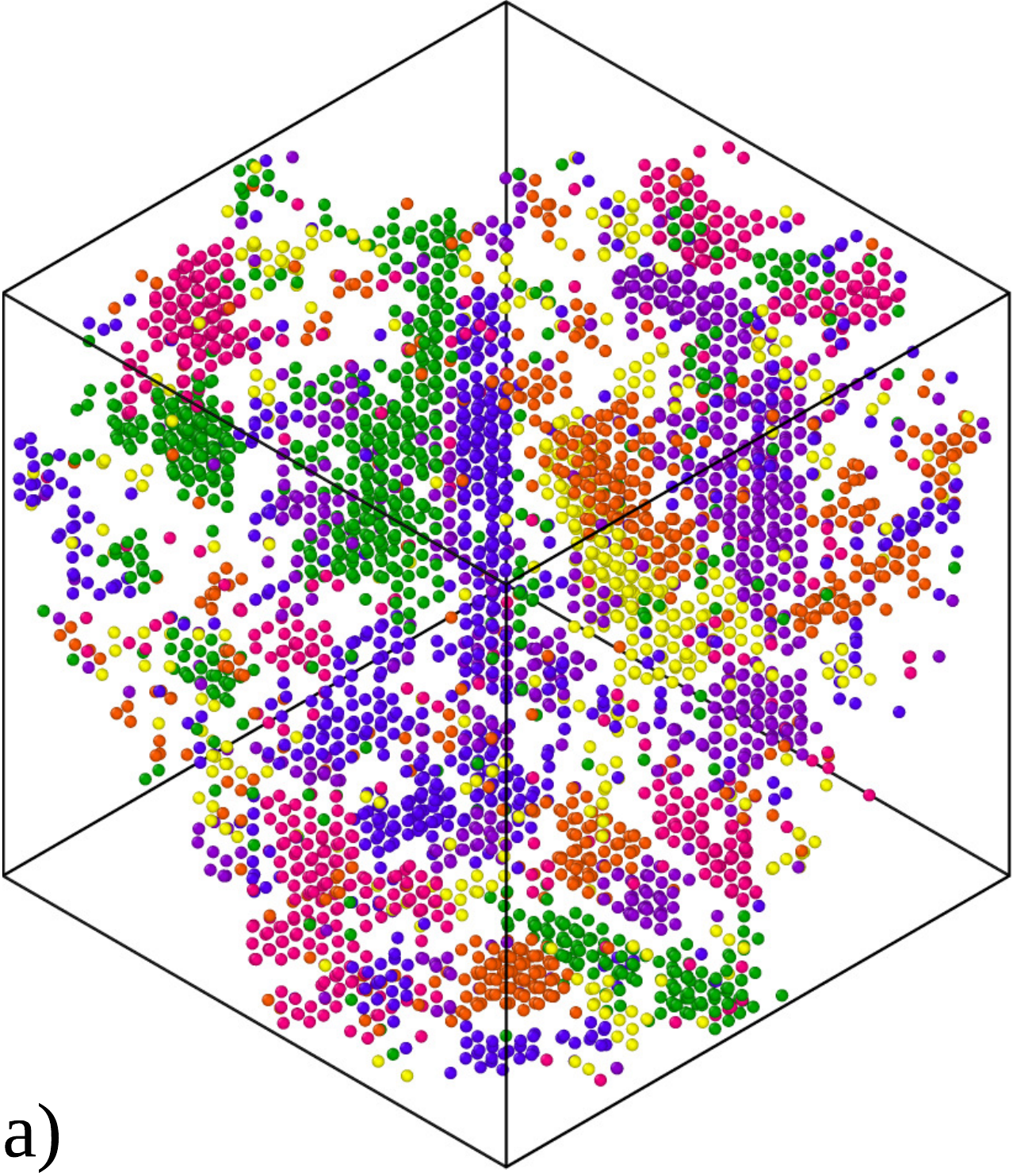}}
	\hspace{0.3cm}
	{\includegraphics[scale=0.23]{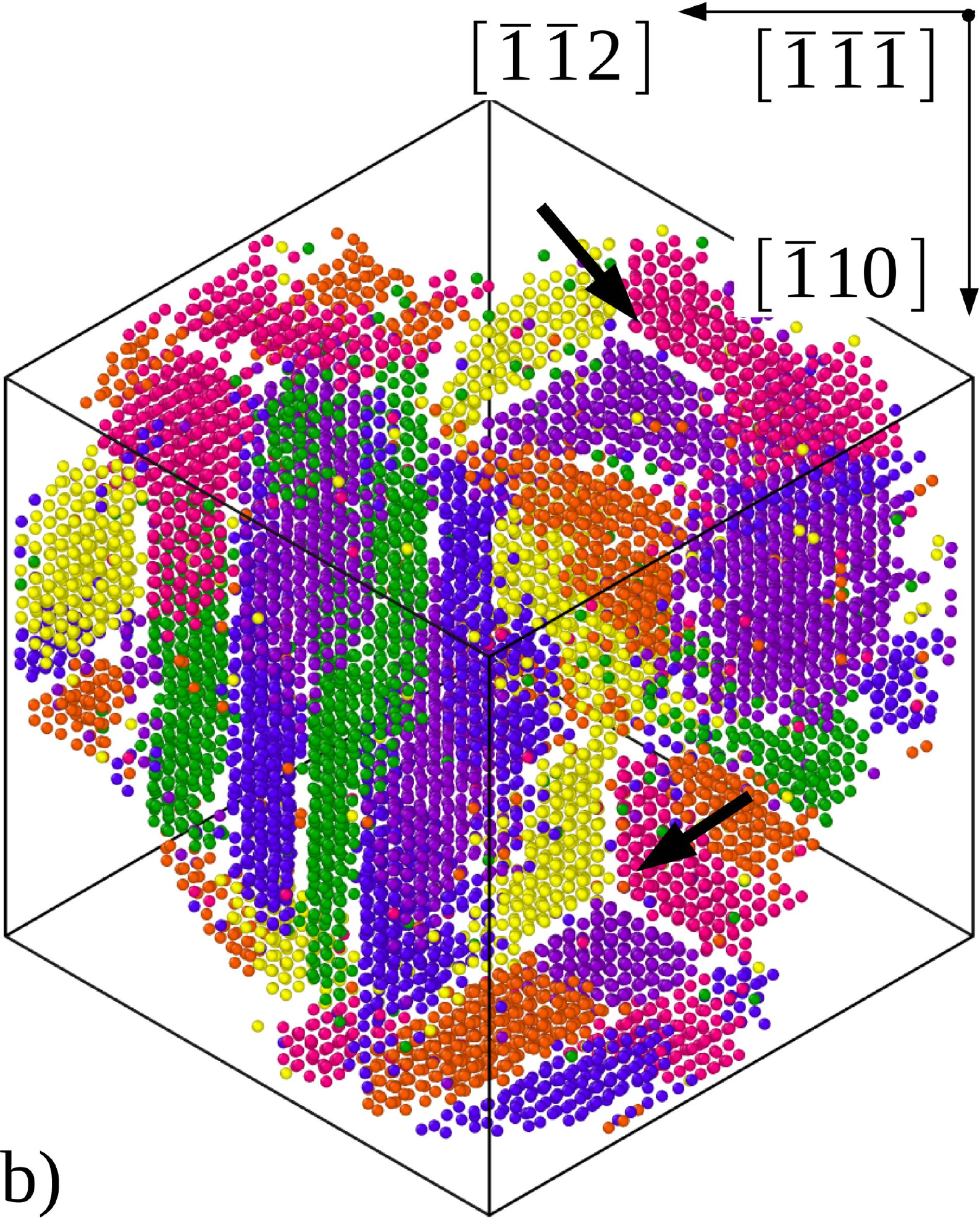}}
	{\includegraphics[scale=0.28]{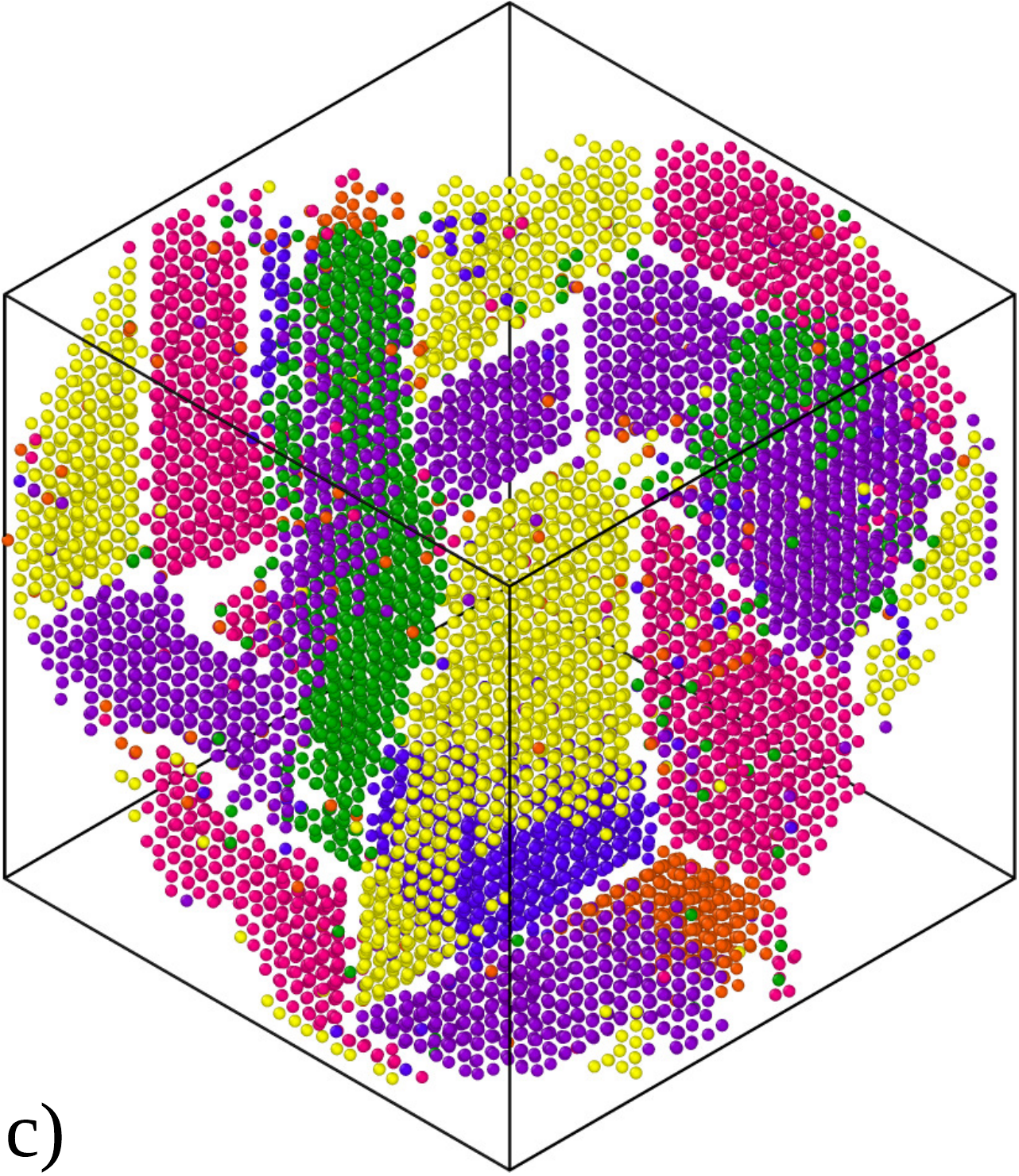}}
	\hspace{0.3cm}
	{\includegraphics[scale=0.28]{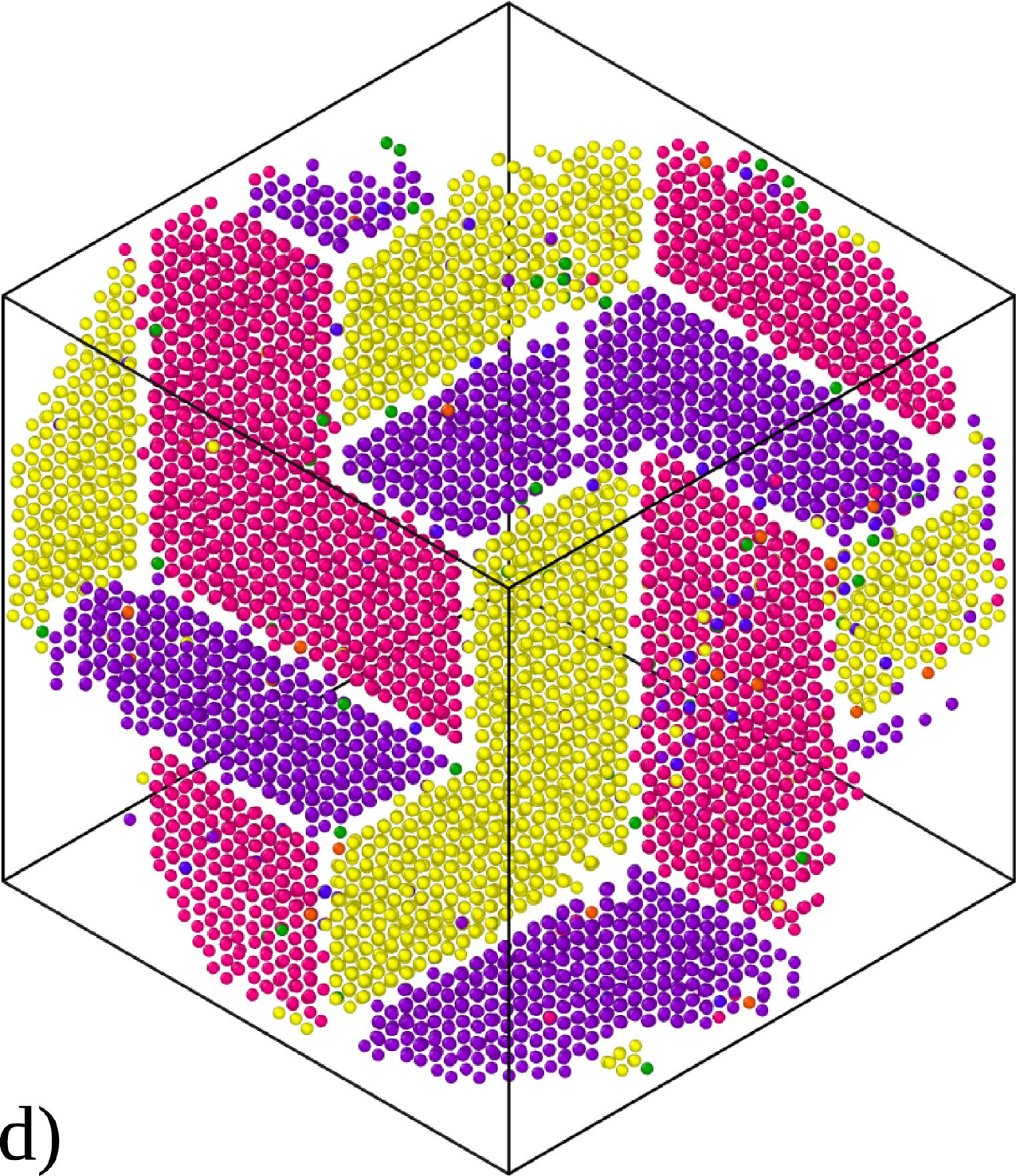}}
	\caption{Evolution of the microstructure in the $(NVT)$ ensemble, only atoms in a small slab normal to the $[111]$ BCC direction and classified as HCP are shown and colored according to the corresponding variant: a) nucleation stage (from A to B in Fig.~\ref{fig:var_sel_4}a), all the variants appear, b) quasi-stationary regime, two stable triple junction, highlighted by arrows, are identifiable, (c-d) the microstructure coarsens (from C to D  in Fig.~\ref{fig:var_sel_4}a) in a 3-plate morphology (final stable state). Crystallographic directions refer to the parent BCC phase.}
	\label{fig:nvt}
\end{figure}
\begin{figure*} [htpb]
	\centering
	\includegraphics[width=0.35\textwidth]{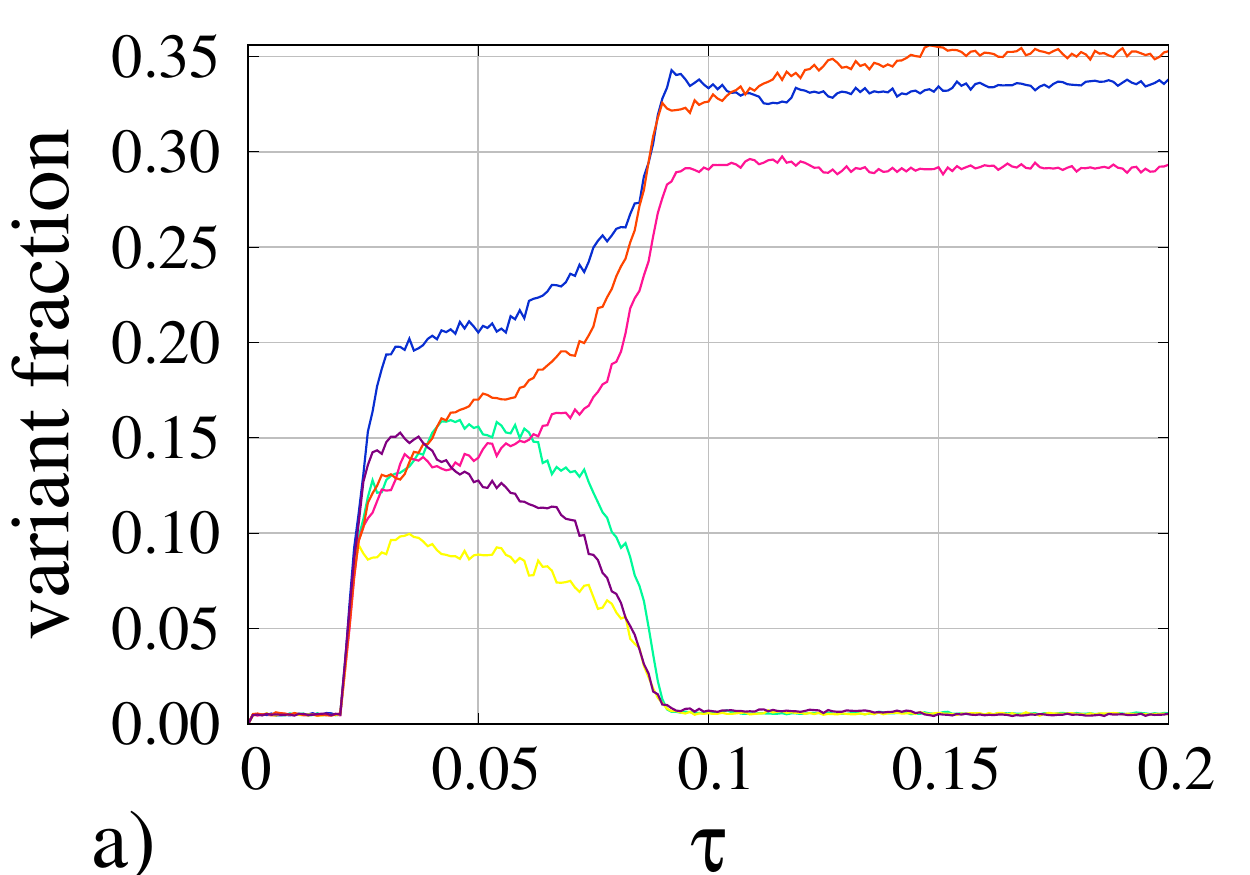}
	\includegraphics[width=0.35\textwidth]{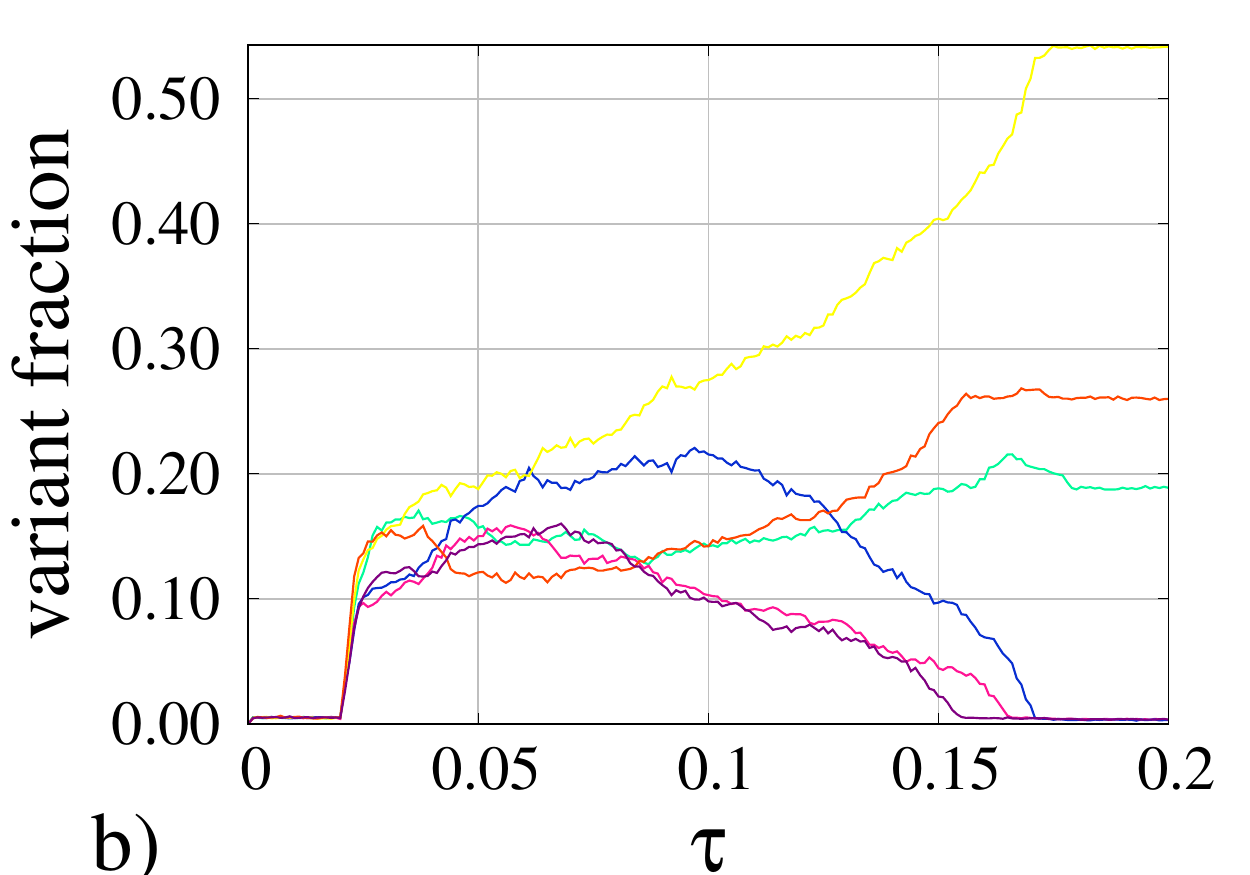}\\
	\includegraphics[width=0.35\textwidth]{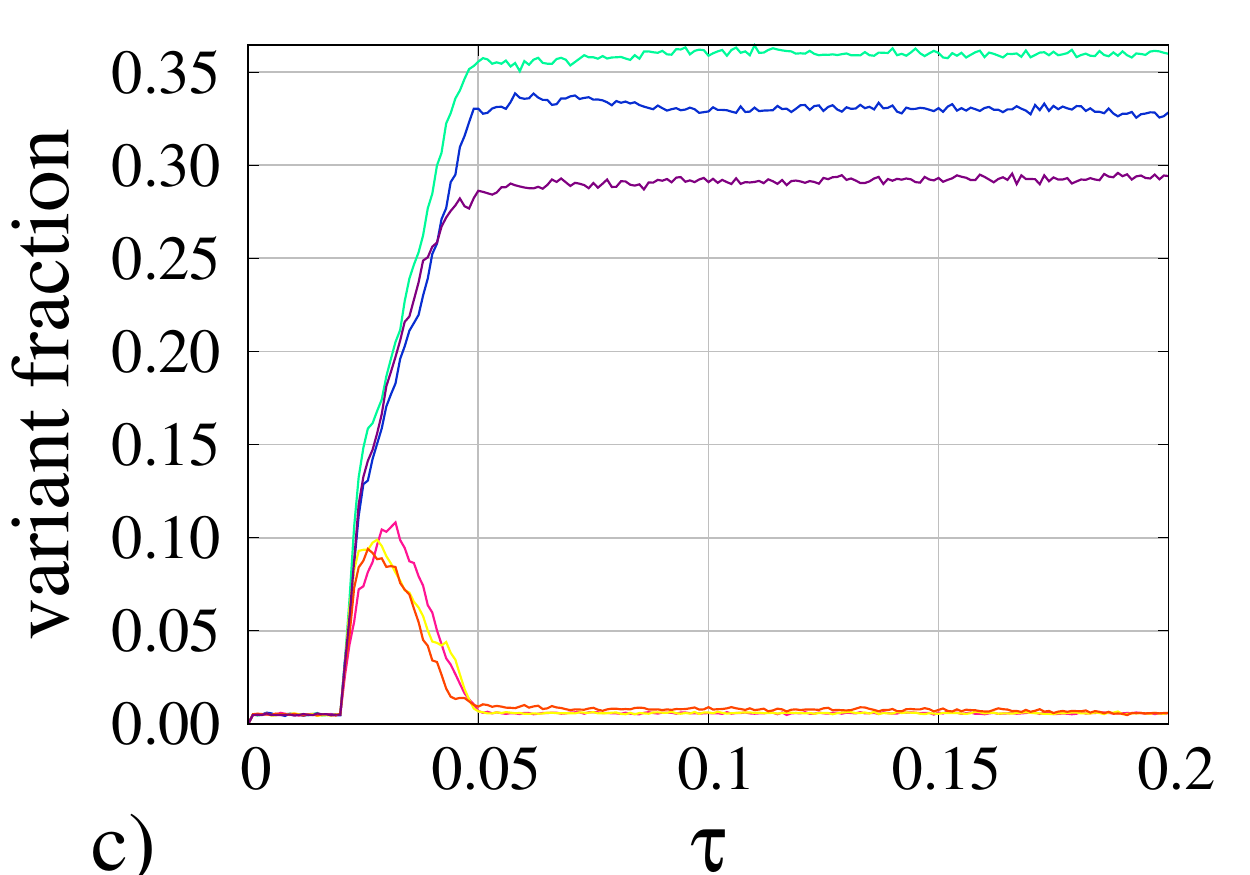}
	\includegraphics[width=0.35\textwidth]{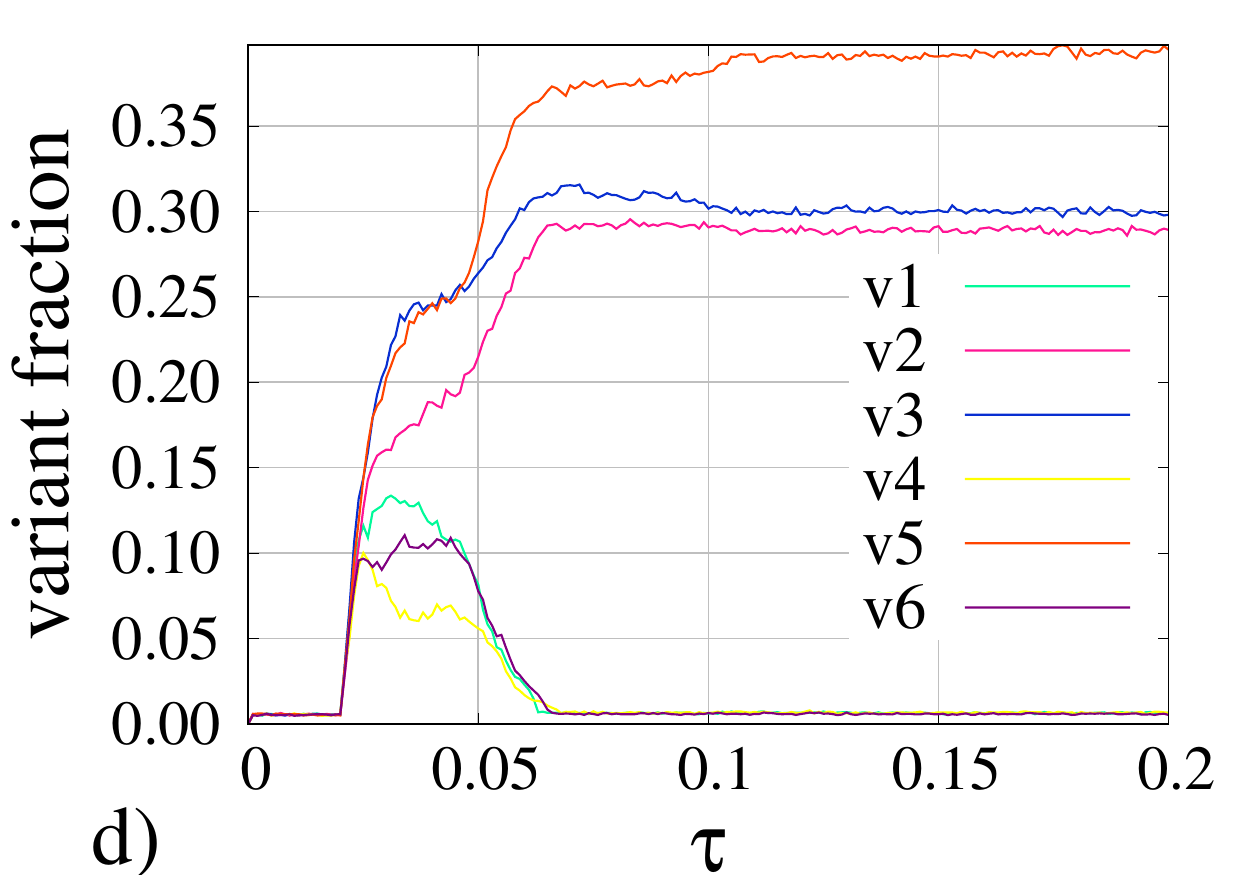}
	\caption{Evolution of the variant fraction in four  different simulations a)-d) in the $(NVT)$ ensemble where the random noise term is different in each case.} 
	\label{fig:fig6}
\end{figure*}
%
%
\section{Discussion}\label{sec:disc}
The results of simulations performed in the $(NVT)$ and $(N\bm{P}T)$ ensembles give highlights on the origin of different defects experimentally observed  in martensite and confirms how deeply local mechanical constraints influence martensite morphology.\\
In both thermodynamic ensembles, at the really beginning of the transition, all variants appear. This is expected, since we consider situations in which all variants are energetically equivalent and therefore equally probable to appear due to the randomness of thermal fluctuations~\cite{gao2014diffuse}. However, when the structure further evolves, only part of them end up to form the final microstructure. We observed two distinct microstructural evolutions, depending on the applied boundary conditions. For simulations in the $(N\bm{P}T)$ ensemble with $\bm{P}= \mathbf{0}$, the microstructure coarsens and forms a single variant domain with antiphase defects where anti-variants domains come into contact (Fig.~\ref{fig:antiphase}). We recall that  a couple of anti-variants are two variants with the same orientation but different shuffling directions \cite{gao2014diffuse}. On the other hand, for simulations in the $(NVT)$ ensemble, a triplet of variants with a common $\langle 11\bar{2}0 \rangle$ HCP direction is systematically selected and form stable triple junctions that drive the overall microstructural evolution (Fig.~\ref{fig:nvt}). In this case, the final microstructure is richer in interfaces and, consequently, has a higher energy with respect to the mono-variant domain obtained for simulations in the $(N\bm{P}T)$ ensemble.\\
As discussed in section \ref{sec:ris}, in almost all the simulations performed in the ($NVT$) ensemble, the 3 selected variants cluster in a 3-plate geometry around the common dense direction. This morphology has been experimentally observed in pure titanium \cite{farabi2018five}, Ti-Nb shape memory alloys \cite{chai2009self}, zirconium alloys \cite{srivastava1993self}, and agrees with predictions based on the phenomenological theory of martensite \cite{bowles1954crystallography}. The 3-plate cluster minimizes the overall mesoscopic shape strain after transition \cite{farabi2018five,srivastava1993self,wang2003effect} because the three selected variants are self-accommodating \cite{miyazaki1989shape,pitteri2002continuum}. Consequently, this morphology is strongly favored when strain energy minimization is dominant in driving the microstructure evolution. The morphology is experimentally observed at different length scales (micrometer \cite{farabi2018five} and sub-micrometer scale \cite{srivastava1993self}) and numerically reproduced by us using a simulation domain of nanometer scale. This suggests that the dominant driving force is the elastic relaxation that largely dominates the interface energy. Furthermore, experiments show how these 3-variant triangles are formed in regions delimited by big martensite laths, which are supposed to originate from first nucleation events. The domain are then progressively filled by smaller and smaller triangles \cite{srivastava1993self,farabi2018five}. These observations together with our simulation results suggest that: i) the 3-variant cluster formation is mainly driven by elastic relaxation, which manifests itself at different length scales. \textcolor{black}{We here showed that, at the nanometer scale, the elastic relaxation is still dominant. Consequently, mechanisms driving the formation of larger microstructures can be easily studied by analysing the evolution of small BCC domains undergoing transition,}  ii)  this 3-variant cluster formation is directly related to martensite nuclei forming under a situation of local confinement. The agreement of our simulations with experiments indicates that simple fixed-volume conditions are well adapted to reproduce this state of local constraints experimented by real systems.\\
We mentioned previously that we occasionally observe the appearance of domains with FCC crystallography. In this case, the 3 variants form two laminates with an  FCC domain at the crossing point. Low-energy coherent interfaces are formed between the HCP $\{0001\}$ basal planes and the $\{111\}$ FCC planes. While in the simulation showing a 3-plate morphology the three variants have similar volume fraction ($\approx 0.3$), in this case one variant (the one participating in both the laminates) is dominant with respect to the others (see Fig.~\ref{fig:var_sel_4}). The possible presence of FCC phase after transition has little experimental evidence \cite{nishiyama1967transmission} and could be an artifact due to the potential. Nevertheless, it has been also reported in other Molecular Dynamics simulations of martensitic phase transition in zirconium \cite{morris2001molecular,pinsook1998simulation,ackland2008molecular}. Subsequent mechanical/thermal treatments, leading to further evolution of the microstructure, could then lead to its progressive extinction in favour of the lower energy HCP phase.\\
\begin{figure} [htpb]
	\centering
	{\includegraphics[scale=0.32]{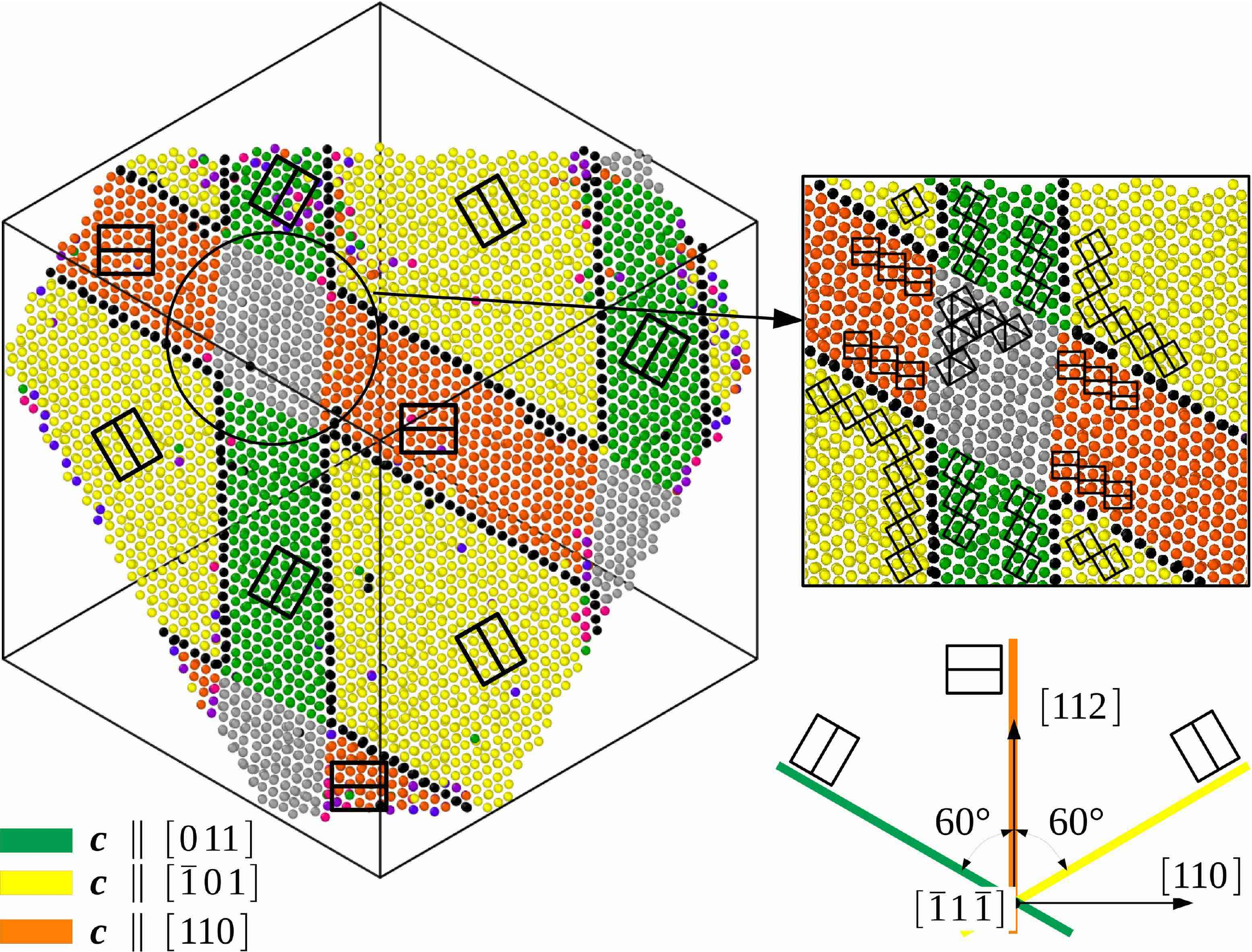}}
	\caption{Final microstructure  at 700 K  in the ($NVT$) ensemble, highlighting the crossing between laminates. Crystallographic directions refer to the parent BCC phase.}
	\label{fig:microstru_nvt_2}
\end{figure}
\begin{figure} [htpb]
	\centering
	\includegraphics[scale=0.33]{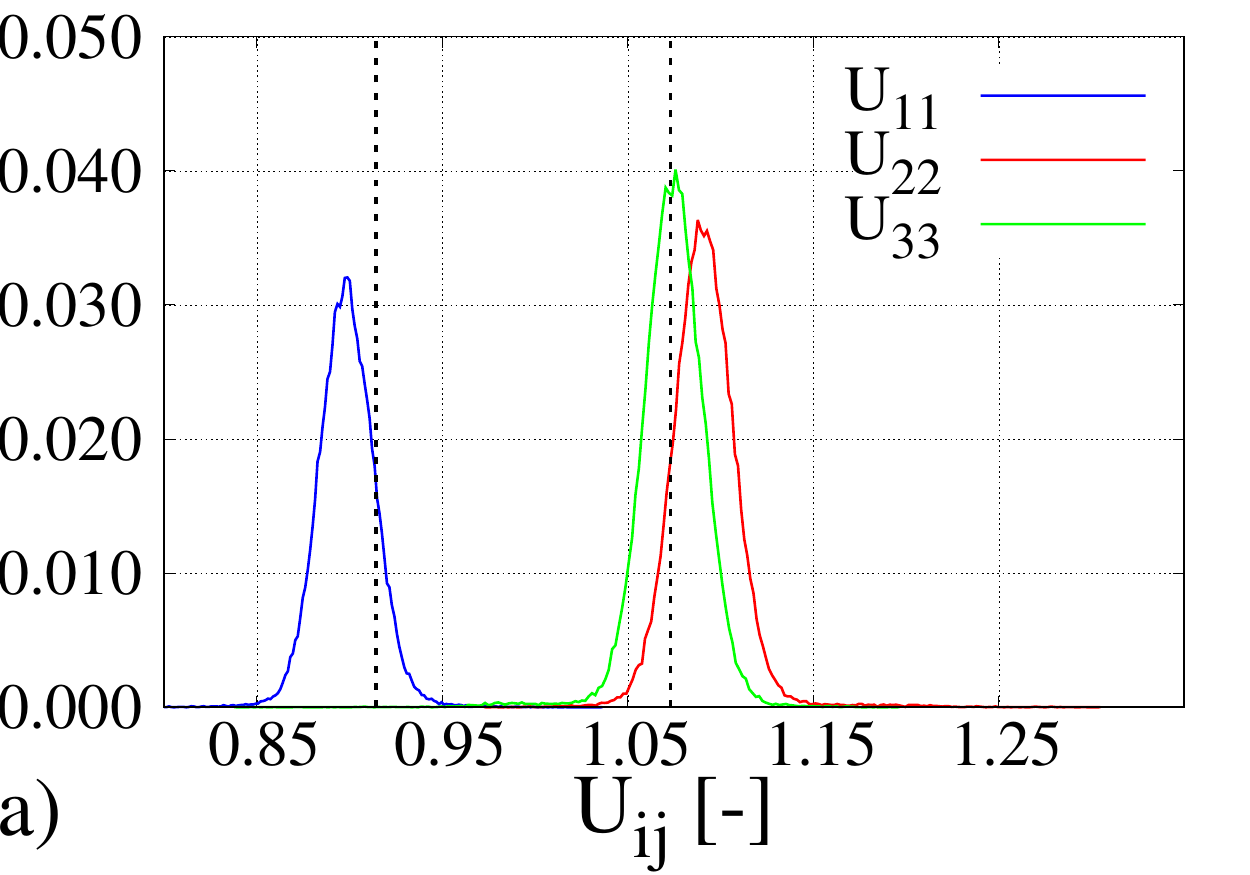}
	\includegraphics[scale=0.33]{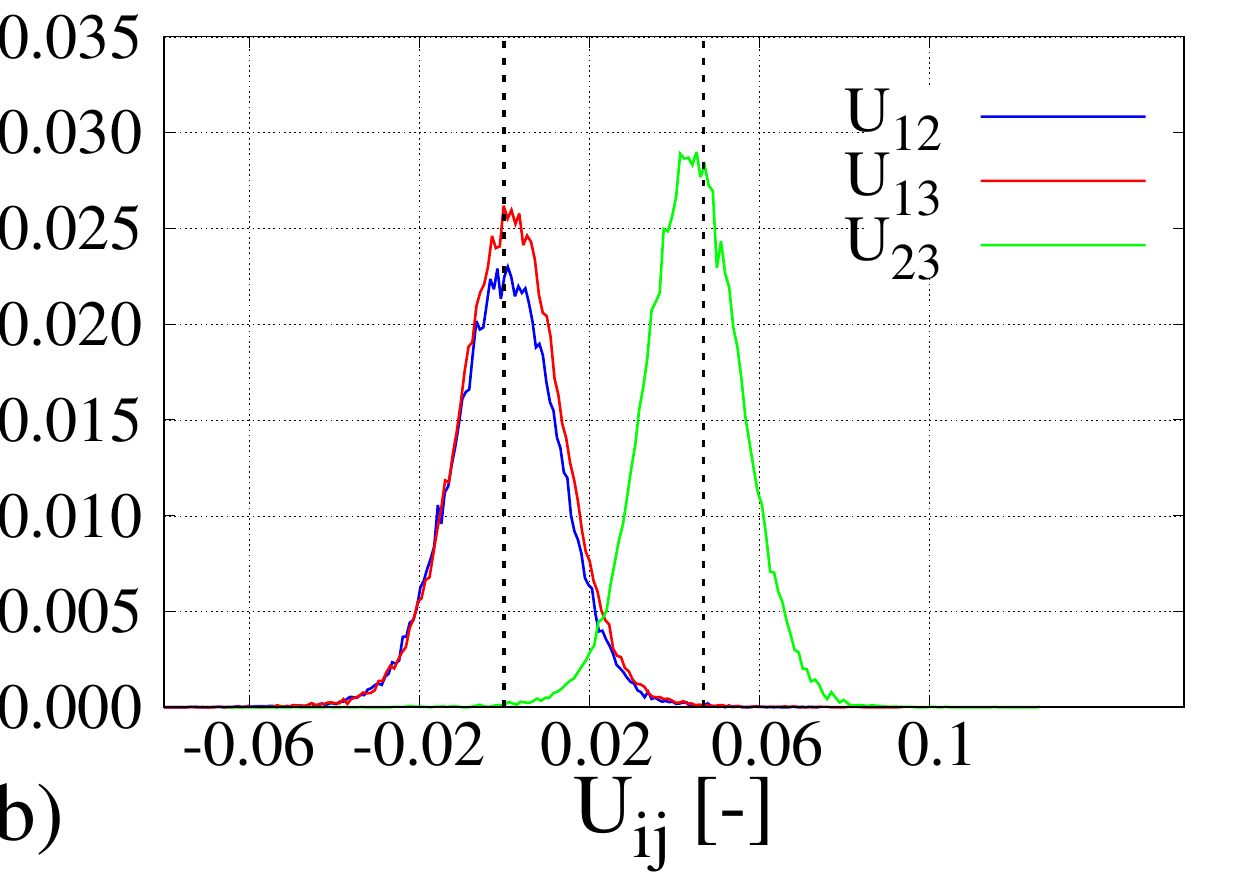}
	\includegraphics[scale=0.33]{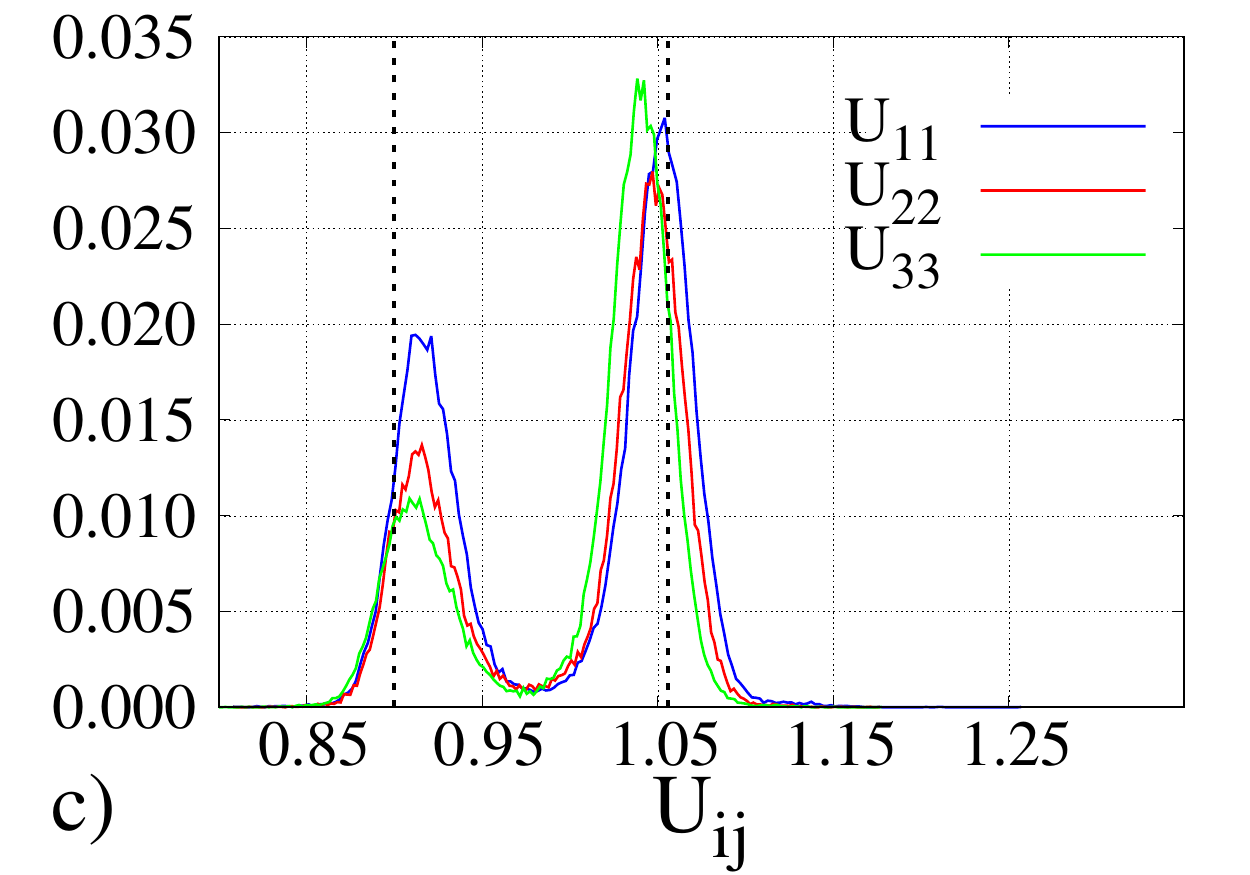}
	\includegraphics[scale=0.33]{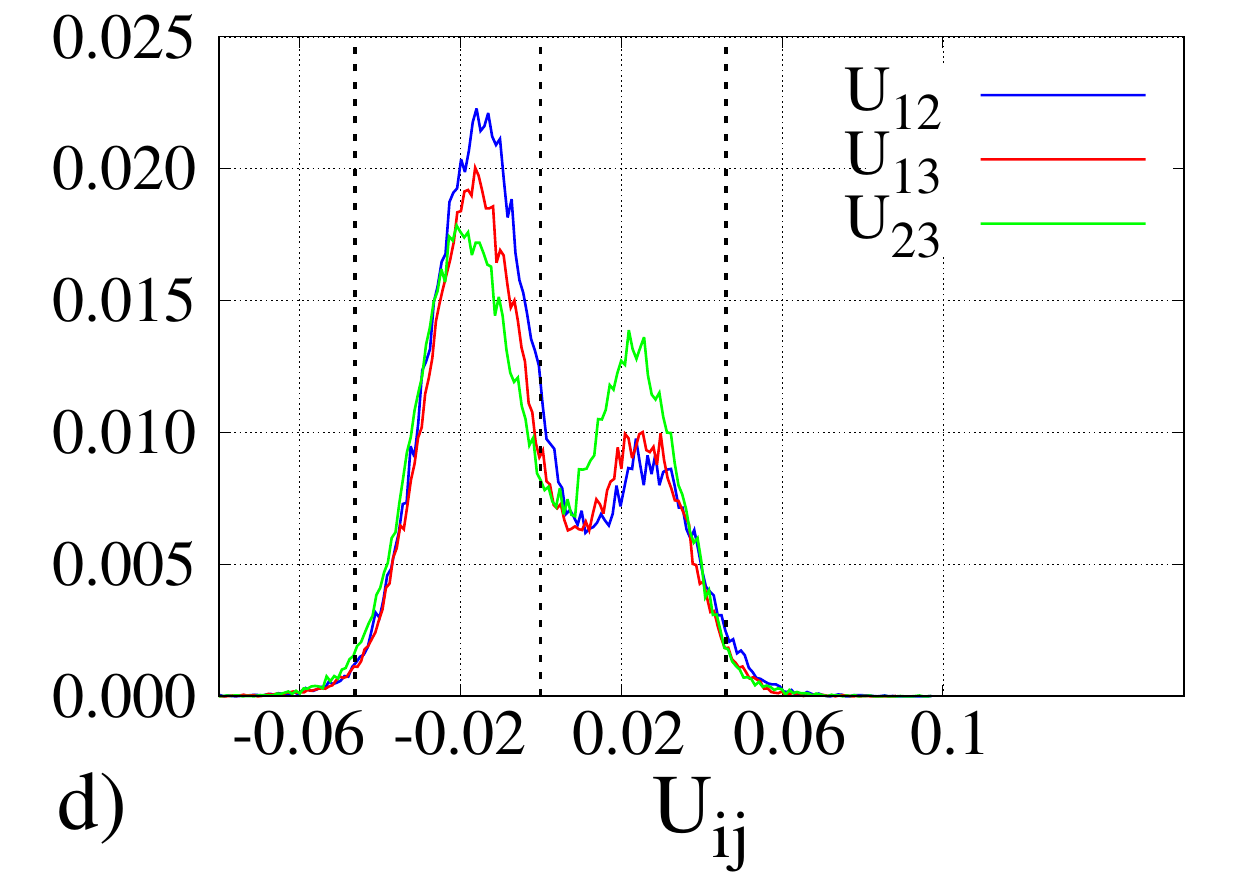}
	\caption{Histograms of $\bm{U}_n$ diagonal and off-diagonal coefficients for a)-b) simulations in the ($N\bm{P}T$) and c)-d) ($NVT$) ensemble. The dotted line shows theoretical values.}
	\label{fig:histo}
\end{figure}
\begin{figure} [htpb]
	\centering
	\includegraphics[scale=0.33]{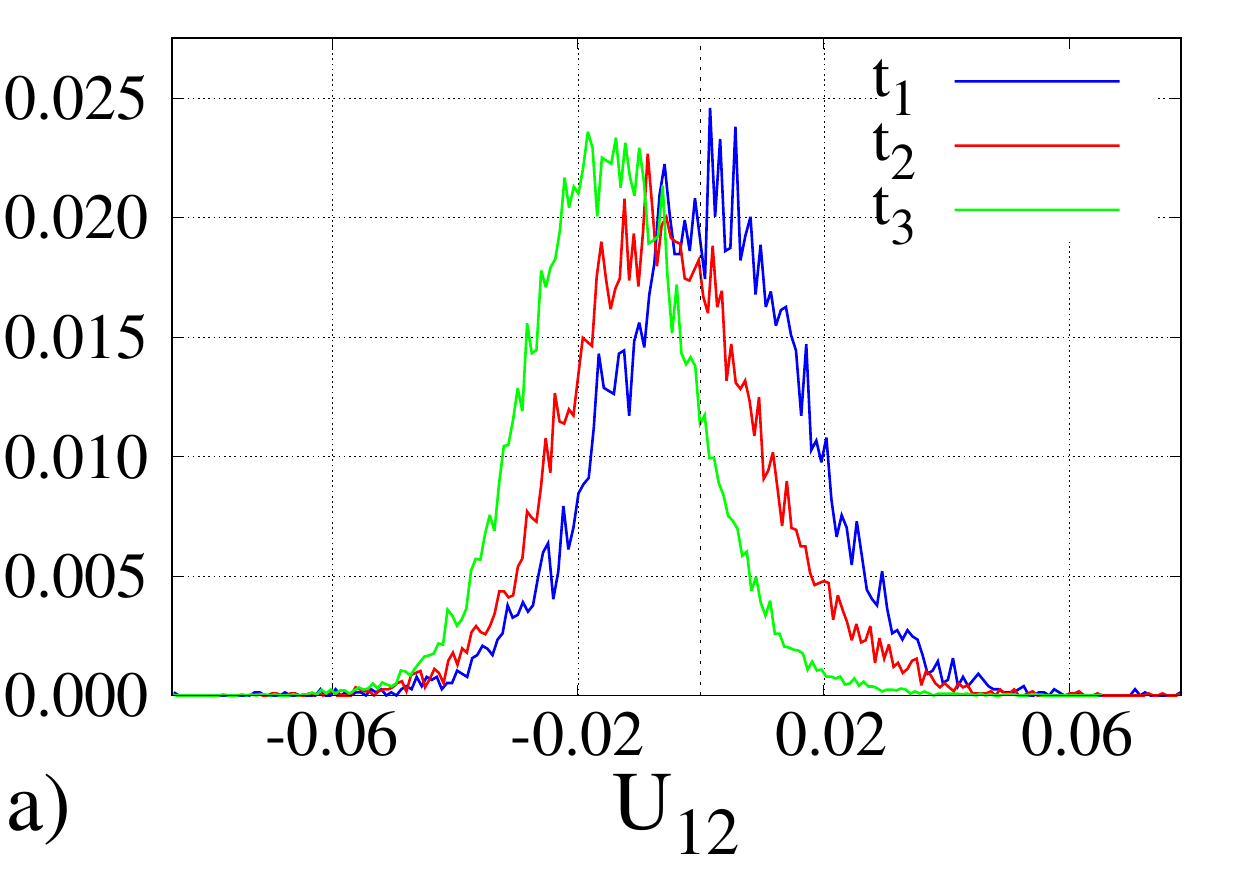}
	\includegraphics[scale=0.33]{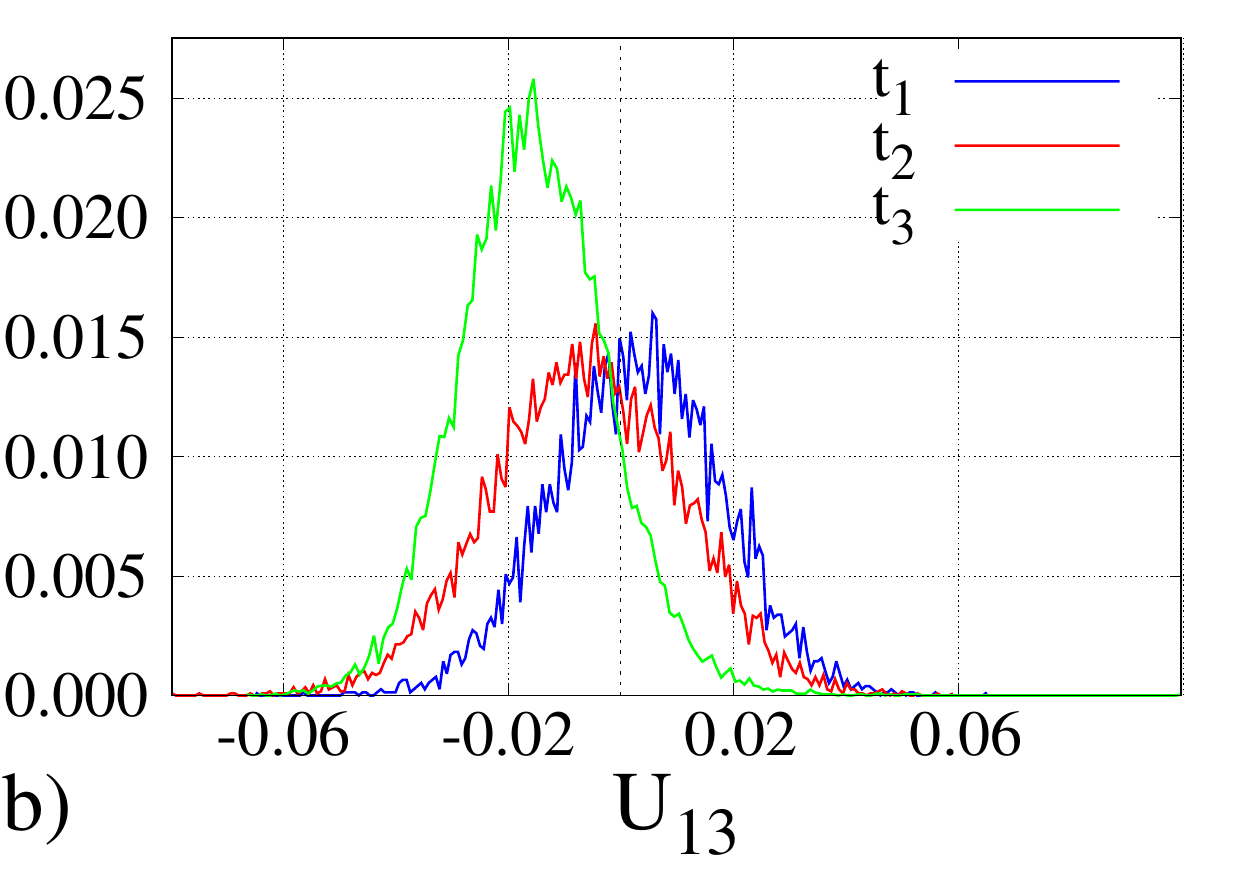}
	\includegraphics[scale=0.33]{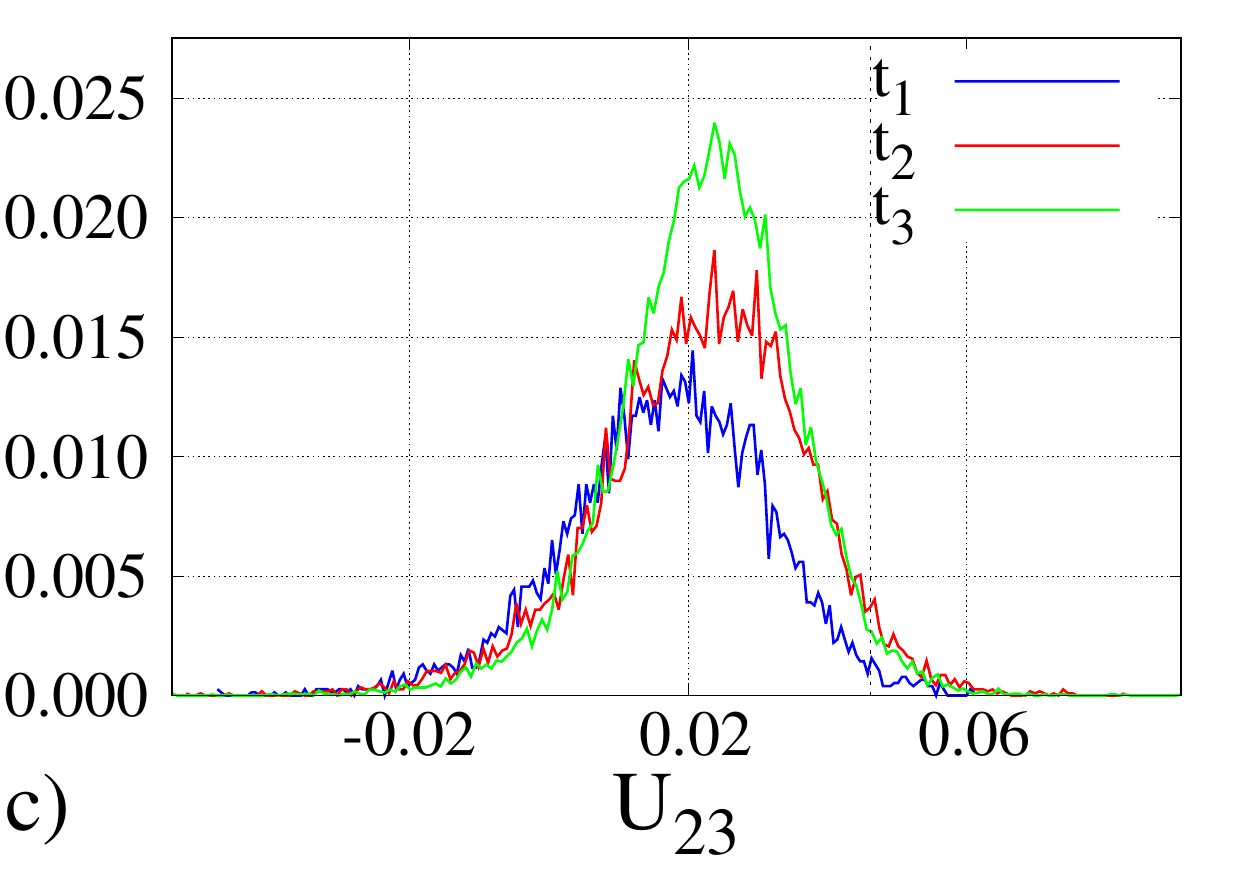}
	\caption{Histograms of off-diagonal coefficients for a single variant $\bm{U}^{(4)}$ in the ($NVT$) ensemble at three different time steps (point A,B,C in the HCP fraction curve of Fig.~\ref{fig:var_sel_4}). Theoretical values are shown in dotted line.}
	\label{fig:histo_2}
\end{figure}
\textcolor{black}{We go now on with the analysis of the defects generated during the transition under different stress conditions. Our simulations give insight on the origin of different kind of interfaces that are found in martensite in pure titanium and corroborates hypothesis based on experimental observations.\\
The single-variant microstructure rising from simulations in the $(N\bm{P}T)$ ensemble (i.e., mimicking the absence of any local constraint in the surroundings) shows antiphase boundaries separating HCP domains with the same orientation of the \textbf{c} axis but different shuffling directions (a couple of anti-variants \cite{gao2014diffuse}). Similarly, single-variant martensite plates containing antiphase boundary networks has been experimentally observed in titanium alloys \cite{banerjee1998substructure} as well as in shape-memory alloys \cite{otsuka2005physical,matsuda2008crystallography}. These early experimental studies led to the hypothesis that these interfaces origin from the nucleation, growth and subsequent impingement of martensite domains, i.e., they are a direct consequence of the randomness of shuffling displacements during the transition \cite{banerjee1998substructure,matsuda2008crystallography}. The microstructural evolution observed in our simulations (see Fig.~\ref{fig:antiphase}), where one variant domain disappear leading two anti-variants domains to come in contact, confirms these hypothesis.\\
For simulations performed in the $(NVT)$ ensemble showing the 3-variants plate morphology, we verified that the grain boundaries separating the different variant domains are all $\{10\bar{1}1\}$ type I twin boundaries (twinning plane $K_1 =\{10\bar{1}1\}$, shear direction $\eta_1 = \langle 1 \bar{2}10\rangle$) \cite{bowles1954crystallography,Bhattacharya2003-lk}), as shown in inset of Fig.~\ref{fig:microstru_1}. This triple junction is not fully consistent with the inter-variant misorientation expected from the Burgers orientation relationship and, from a purely geometrical point of view,  requires some further strain to be present, because the $\{10\bar{1}1\}$ pyramidal plane form an angle of $61.5^\circ$ with the basal HCP plane. The stability showed by these triple junctions during the microstructural evolution (see Fig.~\ref{fig:nvt}) together with the observation of  this specific interface arrangement in all our simulations suggests that the energy cost linked to the additional strain required to form three $\{10\bar{1}1\}$ symmetric boundary is negligible when compared to the energy gain in forming three low energy coherent interfaces \cite{wang2012atomic}. This corroborates experimental observations in pure titanium \cite{wang2003effect,farabi2018five}, titanium alloys \cite{beladi2014variant} and zirconium alloys \cite{srivastava1993self}. In particular, a recent study on grain boundary plane distribution in pure titanium subjected to temperature driven martensitic transformation, which completes previous experimental observations on grain boundary axis angle distribution \cite{wang2003effect}, highlights a strong anisotropy in the grain boundary plane distribution with most of the grain boundaries terminating on $\{10\bar{1}1\}$ pyramidal planes \cite{farabi2018five}. The authors report that these boundaries are associated with symmetric tilt $60^{\circ} [11\bar{2}0]$ inter-variant boundaries and explicitly observed the three-variant cluster in triple junction morphology  \cite{farabi2018five}.}\\
We conclude our discussion by analyzing the numerically computed stretch tensors used to identify variants. Fig.~\ref{fig:histo} shows the histograms of the diagonal and off-diagonal components of $\bm{U}_n$ in the simulations in the ($N\bm{P}T$) and ($NVT$) ensemble, with dotted lines showing the theoretical values predicted by the Mao/Burgers mechanisms. We note that for the simulation in the ($N\bm{P}T$) ensemble, there is an overall good agreement between average values and theoretical predictions. On the other side, in fixed-volume conditions, the deviation is higher and more pronounced in the off-diagonal coefficients. Fig.~\ref{fig:histo_2} compares the histograms of the off-diagonal coefficients for the simulation in the ($NVT$) ensemble at three different time steps $t_1$, $t_2$, $t_3$, corresponding to the end of the nucleation stage, the quasi-stationary regime and the final stationary state (see Fig.~\ref{fig:var_sel_4}). Differently from the histograms of Fig.~\ref{fig:histo}, we report the value for only one of the three variants forming the final microstructure, the $\bm{U}^{(4)}$. The comparison between histograms highlights how the deviation from the theoretical strain values (shown in dotted line) starts developing after the nucleation stage,  when almost all the BCC phase has disappeared and the HCP structure begins coarsening towards the final 3-plate morphology. This is particularly evident for the off-diagonal coefficient $U_{13}$. Although further investigations are needed, we can conclude that these deviations are related to the additional textural evolution arriving in fixed-volume conditions when the first nucleation stage has ended and specific variants domains start growing around the stable triple junctions to form the final 3-plate morphology.
\section{Conclusions}\label{sec:conc}
\textcolor{black}{To summarize, this work is the first numerical study at the atomic scale of the microstructural evolution of pure titanium undergoing temperature-driven BCC$\rightarrow$HCP transition under different stress conditions. For this purpose, we performed a set of extended atomistic simulations with suitable empirical interatomic potential, and we analyzed the microstructural evolution during the transition as well as the final microstructures.}\\
\textcolor{black}{Our main results concern the analysis of the atomistic mechanisms inducing the formation of different defects experimentally observed in martensite in pure titanium and the assessment of the influence of macroscopic constraints on these defects and on the final martensite morphology.\\
When no constraints are present, i.e., when the crystal is allowed to change its shape, a simple mono-variant domain decorated by wavy antiphase boundaries forms. Our simulations confirm previous experimental hypothesis that trace back the origin of these interfaces in the growth and subsequent impingement of variant domains with same orientation but different shuffling directions \cite{banerjee1998substructure,matsuda2008crystallography}.}\\
\textcolor{black}{In contrast, a poly-variant microstructure develops when local constraints prevent a free deformation of the environment surrounding the growing martensite nuclei. This microstructure shows a specific 3-variant morphology which has been extensively documented in experiments on titanium, zirconium and their alloys and which allow the minimization of the strain energy \cite{farabi2018five,srivastava1993self,beladi2014variant}. This microstructure develops around stable triple junctions between variant that are formed at the beginning of the transition after the first nucleation stage. The characterization of interfaces in this microstructure confirms a strong preference for the formation  of boundaries along the $\{10\bar{1}1\}$ HCP pyramidal plane as experimentally documented \cite{farabi2018five}.}\\
\textcolor{black}{Also, we observed the possible appearance of an FCC phase after transition, although further studies are needed to check the absence of any artifact due to the use of empirical inter-atomic potentials \cite{morris2001molecular,pinsook1998simulation,ackland2008molecular}.}\\
 \textcolor{black}{Finally, we stress that this is the first time that the overdamped Langevin dynamics, which has been mostly applied in field of soft matter and bio-molecular simulation, is successfully applied to simulate a fully 3D displacive solid phase transition.This is an important step towards the use of a first-order in time dynamics. The full application of this modelling tool would require its proper derivation through coarse-graining, which will adiabatically eliminate phonons through their incorporation within a coarse-grained potential. This potential will be much softer than the initial one, allowing to use much larger time steps than those required when using the original one.}\\
A natural extension of our work will be to investigate how final microstructures here obtained may influence the mechanical response of the material  under external mechanical loading. This can easily be performed in our formulation through controlling the components of the Piola-Kirchhoff tensor (stress controlled) or the deformation gradient (strain controlled). Our findings can also be useful to develop appropriate mesoscale phase-field theories of BCC-HCP transition, formulated using finite strains  \cite{Vattre2016-ql,Denoual2016-xi} and  Landau-type theories  with  strain components used as the order parameter~\cite{Shchyglo2012-eb}.

\input{appendix_ti_0.tex}

\input{appendix_ti_1.tex}
\input{appendix_ti_2.tex}
\end{document}

%% file: appendix_ti_0.tex
\section*{Appendix A}
We show below that within the statistical ensemble $(N{\bm P}T)$, there is a natural way to define  and compute an internal instantaneous first Piola-Kirchhoff tensor that adopts a virial form and whose statistical average at equilibrium is automatically equal to the externally applied Piola-Kirchhoff stress. We also discuss the link between this instantaneous first Piola-Kirchhoff stress and an instantaneous Cauchy stress, even though the prior computation of a Cauchy stress is not required for our ($N\bm{P}T$) calculations.\\
As explained in the text, within the $(N{\bm P}T)$ ensemble, the internal degrees of freedom associated with the fluctuating box are the entries of the deformation gradient $\bm F$ whose dynamics are given by Eq.~\eqref{eq:lg_ext} which, for the sake of completeness, we recall here
 \begin{linenomath*}
\begin{align}
\frac{d F_{ij}}{dt} & = -  \gamma^{-1} {\frac {\partial \tilde{H}}{\partial F_{ij}}} + \sqrt{2 k_B T \gamma^{-1}} \, \xi_{ij}(t) \ \ \ i,j = 1, ..., 3 \ ,
\label{eq:a1}
\end{align}
\end{linenomath*}
where $ \xi_{ij}(t)$ is a white Gaussian noise such that $ \langle \xi_{ij}(t) \rangle =0$, $ \langle \xi_{ij}(t) \xi_{lm}(t') \rangle =\delta_{ij}\delta_{lm}\delta(t-t')$ and $ \langle \eta_i^n(t) \xi_{lm}(t') \rangle =0$. $\delta_{ij}$ and $\delta_{lm}$ are Kronecker symbols and  $\delta(t-t')$ stands for the Dirac-delta distribution.  As explained in the main text, the driving forces must be computed with the extended Hamiltonian $\tilde{H}$ given by Eq.~(\ref{eq:log})
\begin{linenomath*}
\begin{equation}
\tilde{H}= \Phi(\{F_{ij}L_j^0 \tilde{x}_j^n\}) + V_0 P_{ij} F_{ij} - N k_B T \ln \left( V_0 \det \bm{F}\right),
\label{eq:a2}
\end{equation}
\end{linenomath*}
which must be considered as a function of the scaled coordinates $\{\tilde{x}_i^n\}$ and of the deformation gradient $\bm{F}$. The scaled coordinates are related to the initial atomic coordinates by
\begin{linenomath*}
\begin{equation}
\tilde{x}_i^n = \left( \bm{H}^{-1}\right)_{ij} x^n_j, \ \ \ \ \ \ \ i=1,2,3,
\label{eq:a3}
\end{equation}
\end{linenomath*}
where the matrix $\bm{H}$ is defined by $\bm{H} = \bm{F} \bm{L}^0$, where $ \bm{L}^0$ is a diagonal matrix containing the length of the orthogonal vectors that define the initial simulation box. The driving force that enters Eq.~\eqref{eq:a1} is given by
\begin{equation} 
\begin{split}
 \frac {\partial \tilde{H}}{\partial F_{ij}} &= \frac{\partial \Phi}{\partial F_{ij}}+V_0P_{ij}-N k_B TF_{ij}^{-T}\\
&= \sum_{n,l}\frac{\partial \Phi}{\partial x_l^n}\frac{\partial x_l^n}{F_{ij}}+V_0P_{ij}-Nk_BTF_{ij}^{-T}.
\label{eq:a4}
\end{split}
\end{equation} 
Using Eq.~\eqref{eq:a3}, it is trivial to show that 
\begin{linenomath*}
\begin{equation}
\frac{\partial x_l^n}{\partial F_{ij}}=\delta_{li}x_k^nF_{kj}^{-T},
\label{eq:a5}
\end{equation}
\end{linenomath*}
which inserted in Eq.~\eqref{eq:a4}, leads to
\begin{linenomath*}
\begin{equation}
 \frac {\partial \tilde{H}}{\partial F_{ij}} = \sum_{n}\frac{\partial \Phi}{\partial x_i^n}x_k^nF_{kj}^{-T}+V_0P_{ij}-Nk_BTF_{ij}^{-T}.
 \label{eq:a6}
\end{equation}
\end{linenomath*}
Introducing the virial tensor $\bm{\mathcal{V}}$, defined by
\begin{linenomath*}
\begin{equation}
\mathcal{V}_{ij}=-\sum_n\frac{\partial \Phi}{\partial x_i^n}x_j^n,
\label{eq:a7}
\end{equation}
\end{linenomath*}
the driving forces become
\begin{linenomath*}
\begin{equation}
 \frac {\partial \tilde{H}}{\partial F_{ij}} = -\{Nk_BT\delta_{ik}+\mathcal{V}_{ik}\}F_{kj}^{-T}+V_0P_{ij}.
 \label{eq:a8}
\end{equation}
\end{linenomath*}
We now define the instantaneous internal first Piola-Kirchhoff stress as
\begin{linenomath*}
\begin{equation}
P_{ij}^{inst}=\frac{1}{V_0}(Nk_BT\mathbb{1}+\bm{\mathcal{V}}){\bm F}^{-T},
\label{eq:a9}
\end{equation}
\end{linenomath*}
where $\mathbb{1}$ is the identity matrix. Eq.~\eqref{eq:a8} becomes
\begin{linenomath*}
\begin{equation}
 \frac {\partial \tilde{H}}{\partial F_{ij}} = V_0(P_{ij}-P_{ij}^{inst}),
 \label{eq:a10}
\end{equation}
\end{linenomath*}
The kinetic equation \eqref{eq:a1} then reads
\begin{linenomath*}
\begin{equation}
\frac{d F_{ij}}{ dt}=-\gamma^{-1}V_0(P_{ij}-P_{ij}^{inst})
 +\sqrt{2 k_B T \gamma^{-1}} \, \xi_{ij}(t),
 \label{eq:a11}
 \end{equation}
\end{linenomath*}
At equilibrium, the statistical average of the l.h.s. of this equation is equal to zero (by "statistical average", we mean a time average over a sufficient long time window). As, by definition, the statistical average of the noise term is also equal to zero, we get
\begin{linenomath*}
\begin{equation}
 \langle  P_{ij}^{inst}  \rangle  =P_{ij},
\label{eq:a12}
 \end{equation}
\end{linenomath*}
where $ \langle X \rangle $ stands for the statistical average of $X$. This equation invites us to define an internal first Piola-Kirchhoff stress ${\bm P}^{int}$ as the statistical average of the instantaneous
stress ${\bm P}^{inst}$
\begin{linenomath*}
\begin{equation}
{\bm P}^{int} =  \left \langle {\bm P}^{inst} \right \rangle  = \frac{1}{V_0} \left \langle   (Nk_BT\mathbb{1}+\bm{\mathcal{V}}){\bm F}^{-T} \right \rangle.
\label{eq:a13}
 \end{equation}
\end{linenomath*}
Once the equilibrium is reached, this internal first Piola-Kirchhoff stress, which is defined unambiguously only at equilibrium because it relates on a time average, equilibrates exactly  the imposed
Piola-Kirchhoff stress:
\begin{linenomath*}
\begin{equation}
{\bm P}^{int} =  \bm P.
\label{eq:a14}
 \end{equation}
\end{linenomath*}
We note that the internal stress defined in Eq.~\eqref{eq:a13} adopts a virial form, as it is common to any internal stress defined at the atomistic scale and, therefore, linked to interatomic forces. We also note that the numerical computation of this Piola-Kirchhoff stress is straightforward and does not require the prior computation of any other stress, such as a Cauchy stress, even tough an internal Cauchy stress could also be independently defined and related, through usual relations, to our internal first Piola-Kirchhoff stress (see below). Finally, we also note that this internal first Piola-Kirchhoff stress is not a local quantity defined for each point within the simulation box but rather a global quantity that is associated with the whole system. Therefore, its computation does not require any recipe to define and compute numerically a local stress, such as  the ones proposed by Hardy \cite{hardy1982formulas}.\\

Up to this point, the first Piola-Kirchhoff stress ${\bm P}^{int}$ has been directly defined in the $(N\bm{P}T)$ ensemble within which the deformation gradient ${\bm F}$ is a fluctuating quantity. It has naturally emerged within the kinetic equations associated with the box shape and is such that, at equilibrium, it equilibrates the externally applied  first Piola-Kirchhoff stress. We show now that, as expected, it may also be associated, by conjugacy, with the deformation gradient ${\bm F}$. To show this, we need to introduce the canonical $(N\bm{F}T)$ statistical ensemble, in which $\bm{F}$ is fixed, and to compute the associated free energy $\mathcal{F}$ whose derivative with respect to $\bm{F} $, when properly averaged, will lead to ${\bm P}^{int}$.\\
Within the $(N\bm{F}T)$ ensemble, the partition function $Z$ is given by
\begin{linenomath*}
\begin{equation}
Z=\frac{1}{\Lambda^{3N}N!}\Pi_{n,i}\int dx_i^n e^{-\beta\Phi(\{x_i^n\})},
\label{eq:a15}
 \end{equation}
\end{linenomath*}
where $\Phi(\{x_i^n\})$ is the interatomic potential and $\beta= 1 /(k_B T)$. The term $\Lambda^{3N}$, where $\Lambda$ is the de Broglie wavelength, is reminiscent of the quantum and therefore discrete
nature of the problem. It appears in the classical limit of quantum mechanics and ensures a proper normalization of entropy and free energy. Our aim here is to compute the derivation of the free energy $\mathcal{F}$ with respect to the deformation gradient $\bm{F}$. Therefore, we must introduce the scaled coordinates  $\{\tilde{x}_i^n\}$ which are related to the initial coordinates $\{x_i^n\}$ through the deformation gradient $\bm F$:
\begin{linenomath*}
\begin{equation}
\nonumber
x_i^n=(\bm F \bm L^0)_{ij}\tilde{x}_j^n.
\end{equation}
\end{linenomath*}
The partition function then becomes 
\begin{linenomath*}
\begin{equation}
\nonumber
Z=\frac{1}{\Lambda^{3N}N!}(\det {\bm F \bm L}^0)^N \Pi_{n,i}\int_0^1 d\tilde x_i^n e^{-\beta\Phi( \{\bm F \bm L^0 \tilde{x}^n\})}.
\end{equation}
\end{linenomath*}
Taking the derivative of the free energy $\mathcal{F}=-k_BT \log Z$ with respect to the deformation gradient $\bm F$, we get:
\begin{equation} 
\begin{split}
-\frac{\partial \mathcal{F}}{\partial F_{ij}} & =k_BT \left\{   NF_{ij}^{-T} + \frac{\Pi_{n,i}\int_0^1 d\tilde x_i^n \left(-\beta \frac{\partial \Phi}{\partial F_{ij}}\right)e^{-\beta\Phi}}{\Pi_{n,i}\int_0^1 d\tilde x_i^n e^{-\beta\Phi}}        \right\}\\ & = k_BT \left\{ N F _{ij}^{-T} +  \left \langle -\beta \frac{\partial \Phi}{\partial F_{ij}} \right \rangle _{N{\bm F}T}  \right\}
\\ & =  k_BT \left\{ N F _{ij}^{-T} + \sum_n \left\langle -\beta \frac{\partial \Phi}{\partial x_{i}^n}x_k^n F_{kj}^{-T} \right\rangle _{N{\bm F}T} \right\}.
\end{split}
\label{eq:a16}
\end{equation}
where $\langle  X \rangle_{N\bm{F}T}$ is the statistical average of $X$ within the ensemble ($N\bm{F}T$). We now define the canonical first Piola-Kirchhoff stress as the stress conjugated to the deformation gradient $\bm F$\footnote{Note that the sign convention used here is implied by the enthlapy definition $(\Phi + V_0 \bm{P}\bm{F})$ used to introduce the extended Hamiltonian in Eqs.~\eqref{eq:log} and \eqref{eq:a2}.}
\begin{linenomath*}
\begin{equation}
{\bm P}^{ cano} = - \frac{1}{V_0}\frac{\partial \mathcal{F}}{\partial F_{ij}}.
\label{eq:a17}
\end{equation}
\end{linenomath*}
Using Eq.~\eqref{eq:a16}, we get:
\begin{linenomath*}
\begin{equation}
{\bm P}^{cano} = \frac{1}{V_0} \left \langle \left (Nk_BT\mathbb{1}+\bm{\mathcal{V}}\right){\bm F}^{-T} \right \rangle _{N{\bm F}T},
\label{eq:a18}
\end{equation}
\end{linenomath*}
where, as it does not fluctuate within the $({N{\bm F}T})$ ensemble, the constant term $Nk_BT{\bm F}^{-T}$ has been included within the statistical average and where the virial tensor $\bm{\mathcal{V}}$ has been defined in Eq.~\eqref{eq:a7}.\\
Comparison of Eq.~\eqref{eq:a13} which gives ${\bm P}^{int}$, the internal first Piola-Kirchhoff stress within the $(N{\bm P}T)$ ensemble, and Eq.~\eqref{eq:a18} which gives ${\bm P}^{cano}$, the canonical first Piola-Kirchhoff stress defined within the $({N{\bm F}T})$ ensemble, shows that ${\bm P}^{int}$ and  ${\bm P}^{cano}$ differ only through different statistical averages associated with their respective statistical ensembles. Obviously, for any observable $X$, a statistical average within the $(N{\bm P}T)$  ensemble may be split into a statistical average at a fixed $\bm F$ followed by an average over the fluctuations  of $\bm F$, which, with obvious notations, leads to 
\begin{linenomath*}
\begin{equation}
 \langle X \rangle =\overline { \langle X \rangle }_{N{\bm F}T}^{\bm F}.
\end{equation}
\end{linenomath*}
where, as in Eq.~\ref{eq:a12} and \ref{eq:a13}, $ \langle X \rangle$ refers to the statistical average of $X$ within the $(N\bm{P}T)$ ensemble. Thus, ${\bm P}^{int}$, defined in the $(N{\bm P}T)$ ensemble, is related to ${\bm P}^{cano}$, defined within the $(N{\bm F}T)$  ensemble, by
\begin{linenomath*}
\begin{equation}
{\bm P}^{int}=\overline {{\bm P}^{cano}}^{\bm F}=\overline { \left \langle -\frac{1}{V_0}\frac{\partial \mathcal{F}}{\partial {\bm F}} \right \rangle }_{N{\bm F}T}^{\bm F}.
\end{equation}
\end{linenomath*}
In conclusion, the internal Piola-Kirchhoff stress defined in Eq.~\eqref{eq:a13} is, as expected, related by conjugacy to the deformation gradient ${\bm F}$.\\

The introduction of the first Piola-Kirchhoff stress in the model used here is simply a consequence of the fact that it is the stress measure related by conjugacy to the deformation gradient $\bm{F}$, which, within a lagrangian setup, is the degree of freedom associated to the fluctuating box. Of course, we could also introduce other stress measures, even though this is not needed to integrate our Langevin dynamics. As example, we could define an instantaneous Cauchy stress in such a way that it is related to the instantaneous first Piola-Kirchhoff stress through the usual relation:
\begin{linenomath*}
	\begin{equation}
\bm{\sigma}_1^{inst} = \frac{1}{\det \bm{F}} \bm{P}^{inst} \bm{F}^T,
\label{eq:a21}
	\end{equation}
\end{linenomath*}
which, with Eq.~\eqref{eq:a9}, leads to:
\begin{linenomath*}
	\begin{equation}
	\bm{\sigma}_1^{inst} = \frac{1}{V}(Nk_BT\mathbb{1}+\bm{\mathcal{V}}),
	\label{eq:a22}
	\end{equation}
\end{linenomath*}
whose statistical average in the ($N \bm{P} T$) ensemble leads to the definition of an internal Cauchy stress:
\begin{linenomath*}
	\begin{equation}
	\bm{\sigma}_1^{int} = \left \langle \frac{1}{V}(Nk_BT\mathbb{1}+\bm{\mathcal{V}}) \right \rangle.
	\label{eq:a23}
	\end{equation}
\end{linenomath*}
The question now arises as to whether this internal Cauchy stress is equal, at equilibrium, to an external Cauchy stress. Naturally, we would like this external Cauchy stress $\bm{\sigma}$ to be related to the applied first Piola-Kirchhoff stress by the usual relation:
\begin{linenomath*}
	\begin{equation}
\bm{\sigma} = \frac{1}{\langle \det \bm{F} \rangle} \bm{P} \langle \bm{F} \rangle^T.
	\label{eq:a24}
	\end{equation}
\end{linenomath*}
Using Eqs.~\eqref{eq:a13} and \eqref{eq:a14}, which say that, at equilibrium in the $(N \bm{P} T)$ ensemble, the average of the instantaneous first Piola-Kirchhoff stress is equal to the applied first Piola-Kirchhoff stress, and Eq.~\eqref{eq:a21} for the definition of the instantaneous Cauchy stress, we get:
\begin{linenomath*}
	\begin{equation}
	\bm{\sigma} = \frac{1}{\langle \det \bm{F} \rangle} \langle \det \bm{F} \bm{\sigma}_1^{inst} \bm{F}^{-T} \rangle \langle \bm{F}\rangle^T.
	\label{eq:a25}
	\end{equation}
\end{linenomath*}
Because of the coupling between the fluctuations of $\det \bm{F}$, $\bm{\sigma}^{inst}$ and $\bm{F}$, the r.h.s of this equation cannot be further simplified. Therefore, strictly speaking, the statistical average of the instantaneous Cauchy stress as defined in Eqs.~\eqref{eq:a21}-\eqref{eq:a22} is not equal to the applied Cauchy stress defined in Eq.~\eqref{eq:a24}
\begin{linenomath*}
	\begin{equation}
	\bm{\sigma} \ne  \langle \bm{\sigma}_1^{inst}\rangle.
	\label{eq:a26}
	\end{equation}
\end{linenomath*}
As an alternative, we could define an instantaneous Cauchy stress in the following way:
\begin{linenomath*}
	\begin{equation}
	\bm{\sigma}_2^{inst} = \frac{1}{\langle \det \bm{F} \rangle} \bm{P}^{inst} \langle \bm{F} \rangle^T.
	\label{eq:a27}
	\end{equation}
\end{linenomath*}
Using Eqs.~\eqref{eq:a13}, \eqref{eq:a14} and \eqref{eq:a24} we immediately see that the statistical average of this instantaneous stress is now equal to the applied Cauchy stress:
\begin{linenomath*}
	\begin{equation}
	\bm{\sigma} =  \langle \bm{\sigma}_2^{inst}\rangle.
	\label{eq:a28}
	\end{equation}
\end{linenomath*}
However, we note that, strictly speaking, the definition of the instantaneous Cauchy stress given in Eq.~\eqref{eq:a27} is not entirely satisfactorily because its numerical application requires the prior knowledge of the statistical averages of $\det \bm{F}$ and $\bm{F}$. As a final comment, we note that in the thermodynamic limit of an infinite system, fluctuations may be neglected (provided the system is not going through a phase transition): definitions \eqref{eq:a21} and \eqref{eq:a27} become equivalent and Eq.~\eqref{eq:a26} becomes and equality.

%% file: appendix_ti_1.tex
\section*{Appendix B: integration scheme}
Our Langevin dynamics is defined by stochastic equations with white noise (see Eqs.~\eqref{eq:lg_ext}) that display the following generic form:
\begin{linenomath*}
\begin{equation}
\text{d}X_i(t) = a_i(\{X_i\}) dt + B \text{d}W_i(t)
\label{eq:app1}
\end{equation}
\end{linenomath*}
where $ a(\{X_i\})$ is a drift term, $B$ a noise amplitude and the differential $dW(t)$ denotes an infinitesimal increment of the Wiener process $W_i(t)$. The integral form of Eq.~(\ref{eq:app1}) is:
\begin{linenomath*}
\begin{equation}
X_i(t) = \int_{0}^{t}a_i(\{ X_i(t)\}) \text{d}t + \\
\int_{0}^{t} B\text{d}W_i(t) + X_i(t_0),
\label{eq:app2}
\end{equation}
\end{linenomath*}
where the first term in the r.h.s. is a Riemann integral and the second term is a stochastic integral. To numerically evaluate Eq.~(\ref{eq:app2}) we used an explicit predictor-corrector method which results in the following numerical scheme \cite{Burrage2007-gj,baruffi2019overdamped}:
\begin{linenomath*}
\begin{equation} \label{eq:app3}
\begin{split}
X_i& (t + \Delta t) = X_i(t) + \\
&+ {\frac{\left[a_i(\{ \bar X_i(t+\Delta t)\})+a_i(\{X_i(t)\})\right]}{2}} \Delta t  \\ & +   B \Delta W_i(t),
\end{split}
\end{equation}
\end{linenomath*}
where the finite increment $\Delta W_i(t) = W_i(t + \Delta t) - W_i(t)$ can be calculated as $\Delta W_i(t) = \sqrt{\Delta t} \xi(t)$ with $\Delta t \in \mathbb{R}$ and $\xi(t)$ taken from a normal distribution with unit variance. $\bar X_i$ is evaluated using the explicit Euler method:
\begin{linenomath*}
\begin{equation}
\bar X_i(t + \Delta t)= X_i(t) +a_i(\{X_i(t)\}) \Delta t + B\Delta W_i(t).
\end{equation}
\end{linenomath*}

%% file: appendix_ti_2.tex
\section*{Appendix C: numerical procedure for variant identification}
\begin{figure} [htpb]
	\centering
		\includegraphics[width=0.45\textwidth]{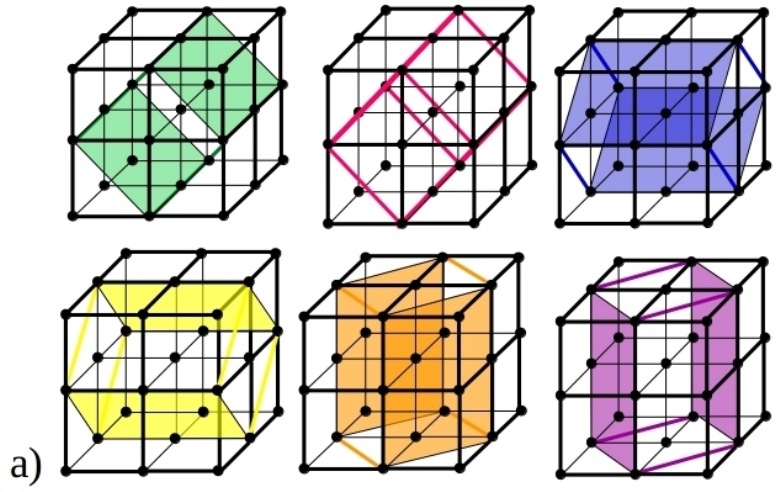}
		\includegraphics[width=0.22\textwidth]{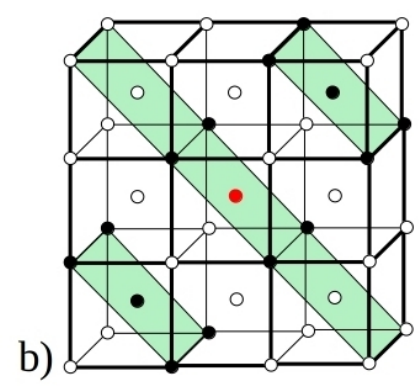}
	\caption{a) 
	\textcolor{black}{Six possible orientations of the (110)$_{bcc}$ planes which may transform into (0001)$_{hcp}$ planes during the BCC-HCP transformation.
Note that the central atom of the BCC cubic cell is not shown for the sake of clarity. b) Example of a neighbor set $\Omega_n$ (colored in black, containing 14 atoms) for a given atom $n$ (colored in red) for one of the six orientations.}}
	\label{fig:basal_planes}
\end{figure}

The numerical procedure used to identify the different variants is based on the definition of a local strain describing the transition from the BCC to the HCP structure. To calculate this local atomic strain, we implemented the following procedure as described in \cite{wuvariant}.\\
We start from a BCC structure with crystal axis $\langle 100 \rangle_{BCC}$ parallel to the mainframe axis. For every atom \textit{n}, we consider six possible sets of 14 neighbors by taking six different configurations defined based on the six cubic cells which can deform into the orthorhombic one, as schematically illustrated in Fig.~\ref{fig:basal_planes}. As already mentioned, the transformation from BCC to HCP cannot be fully described by a simple homogeneous deformation gradient, but supplementary atomic displacements applied on a sublattice of the deformed lattice are needed. These displacements consist in an alternate shuffling of $\{110\}$ planes along the $\langle \bar{1}10 \rangle$ directions. The overall deformation of the lattice does not describe the shuffling so, to numerically evaluate the local strain, half of the atoms of the original BCC lattice must be considered in the six possible configurations.\\
After defining the six possible sets of neighbors, we calculate for each atom \textit{n} six deformation gradients $\bm{F}^{(k)}_n$, following the approach proposed by Falk \cite{falk1998dynamics} and, by polar decomposition, six strains $\bm{U}^{(k)}_n$, each one associated to a $\{110\}_{BCC}$ plane in the undeformed configuration. The local strain for atom $n$ is then defined as the $\bm{U}^{(k)}_n$ with minimum $D_n^{{(k)}^2}$. Based on this assignment, we label the atom $n$ as belonging to the corresponding variant.
